\documentclass{aa}

\usepackage{natbib}
\usepackage{graphicx}
\usepackage{epstopdf}
\usepackage{graphics}
\usepackage{rotating}
\usepackage{color}

\bibpunct{(}{)}{;}{a}{}{,}

\begin{document}

\def\HI{H{\sc i} }
\def\HII{H{\sc ii} }
\def\Ha{{\rm H}\alpha }
\def\Msun{M_{\odot} }
\def\kms{\,km\,s^{-1}}

\title{Do High-Velocity Clouds trace the Dark Matter subhalo population?}
\titlerunning{Do HVCs trace the DM subhalo population?}

\author {S.~Pl\"ockinger \and G.~Hensler}
\institute{University of Vienna, Institute of Astrophysics, Tuerkenschanzstr. 17, 1180 Vienna, Austria; \\
email: sylvia.ploeckinger@univie.ac.at, gerhard.hensler@univie.ac.at}

\abstract{
Within the cosmological concordance model, Cold Dark Matter (CDM) subhalos form the building blocks which merge hierarchically to more massive galaxies. This concept requires that even nowadays massive galaxies are surrounded by numerous subhalos. Since intergalactic gas is accreted by massive galaxies, observable e.g. as high-velocity clouds (HVCs) around the Milky Way, with extremely low metallicities, these can be suggested to represent the baryonic content of primordial Dark Matter (DM) subhalos. Another possibility of their origin is that they stem from disrupted satellite galaxies, but in this case, these gas clouds move unaccompanied by a bound DM structure.   }
{HVCs are mainly observed at significant distances from the galactic disk, while those closer to the disk have on average lower infall speed and larger but not yet solar element abundances. This can be caused by interactions with the hot halo gas: decelerated by ram pressure and metal-enriched by the more abundant gas in the galactic halo. Since HVCs are observed with long gas tails and with irregular substructures, numerical models of gas clouds passing through hot halo gas are performed aiming at exploring their structure and compare them with observations. If HVCs are engulfed by DM subhalos, their gas must leave the DM gravitational potential and reflect this in their dynamics. On the other hand, the evolution and survival of pure gas models must be tested to distinguish between DM-dominated and DM-free clouds and to allow conclusions on their origin. }
{A series of high-resolution hydrodynamical simulations using the adaptive-mesh refinement code FLASH V3.2 are performed for typical HVC masses, distances, and infall velocities. }
{The models demonstrate that purely baryonic HVCs with low masses are disrupted by ram-pressure stripping and Kelvin-Helmholtz instabilities, while more massive ones survive, losing their initially spherical shape and develop significant substructures including cometary elongations in the column density distribution (``head-tail structure"). On the contrary, HVCs with DM subhalos survive with more than 90\% of their gas mass still bound and spherically shaped, approaching the Galactic disk like bullets.
In addition, we find that velocity gradients along the cometary head-tail structures does not necessarily offer a possibility to distinguish between DM-dominated and purely gaseous HVCs. Comparison of models with observations let us conclude that HVCs are not embedded in a DM substructure and do not trace the cosmological subhalo population.}
{}


\maketitle

\section{Introduction}

In his general investigation of a possible Galactic corona, \citet{1956ApJ...124...20S} was the first who predicted the existence of high-velocity clouds (HVCs). A few years later the systematic search for neutral hydrogen clouds succeeded and E. Raimond observed the 21 cm line of neutral hydrogen at high velocities \citep{1963Natur.200..155M}.

Although the possibilities in observations, the theoretical understanding of the interstellar medium (ISM), and the facilities for numerical simulations have increased drastically, yet half a century later, the formation and evolution of HVCs are still under controversal discussion. The theories on their origin do not only vary in small details but lead to completely different pictures of the observed clouds, with tremendous consequences for galaxy formation and evolution. They are approaching the Galactic disk with high velocities and, depending on their distance, have to pass the hot coronal gas of the Milky Way halo.

Their radial velocities lie above or around escape velocity of the Milky Way and 
their rotation component is incompatible with the Galactic rotation \citep{1999ApJ...514..818B}.
This obtrudes the extragalactic origin of these gas clouds, excluding the 
widely discussed possibility of  ejecta from galactic fountains 
\citep{1976ApJ...205..762S, 1980ApJ...236..577B}.

This issue leaves open two further fundamental possibilities: 
They are: 1) relics of cosmological structure formation in a Cold Dark Matter (CDM) cosmology 
\citep{2006ApJ...646L..53C}
and therefore distant ``building blocks" from galaxy formation in the 
Local Group \citep{1999ApJ...514..818B} 
or 2) remnants from galaxy collisions, from tidal disruption of dwarf galaxies \citep{2004ApJ...603L..77P}, 
or/and from close-by tidal interactions of satellite galaxies
\citep{1996MNRAS.278..191G, 1998MNRAS.299..611B}. 
The main difference to distinguish between both sources is their 
Dark Matter (DM) content: in case 1) clouds should be strongly dominated by DM, but not in 2). 
Decisive issues could be derived from detailed observations in connection with complex numerical simulations, accounting for their distances, physical state, dynamical structure and chemical abundances.

\begin{table*}[htdp]
\caption{Distances and Abundances of HVC and IVC}
\begin{center}
\begin{tabular}{lllc}
\hline
\hline
Cloud 		&		Distance 	&	Abundance		& References \\
			&		(kpc)		&	(rel. to solar)	&	\\
\hline
Complex A	&		4.0 - 9.9 	&	0.02 - 0.4			& 1\\
Complex C 	&		$10 \pm 2.5$	&	0.09 - 0.29	& 2,3  \\
Complex H 	&		$27 \pm 9$&					& 4 \\
CHVC 224.0-83.4-197&	?		&	$< 0.46$			& 5 \\		
CHVC 125+41-207&	?		&	$<0.2$			& 6 \\
Upper IV Arch	&		0.8 - 1.8 	&	$\approx 1$		& 1\\
Lower IV Arch 	&		0.4 - 1.9	&	$\approx 1$ 		& 1\\
LL IV Arch 	&		0.9 - 1.8 	&	$1.0 \pm 0.5$ 		& 1\\ 
Complex K	&		$< 6.8$	&	$< 2.0$			& 1\\
Pegasus-Pisces Arch &	$< 1.1$	&	$0.54 \pm 0.04$	& 1\\
Complex gp	&		$0.8-4.3$	&	1-2				& 1\\	
\hline
\end{tabular}
\tablebib{
(1)~\citet{2001ApJS..136..463W}; (2)~\citet{2008ApJ...684..364T}; (3)~\citet{2007ApJ...657..271C}; 
(4)~\citet{2003ApJ...591L..33L}; (5)~\citet{2002ApJ...572..178S}; (6)~\citet{2001A&A...370L..26B}.
}
\end{center}
\label{tab:metall}
\end{table*}%

\subsection{Observations}

\subsubsection{Distances, All-Sky Surveys, General Properties}

Detailed studies measure distances of a few kpc for the extended HVC complexes \citep{2006ApJ...638L..97T, 2008ApJ...672..298W}, e.g. Complex C with a distance of $d = 10\pm2.5\, \mbox{kpc}$ \citep{2008ApJ...684..364T}. For the massive Complex H a distance of 27 kpc \citep{2003ApJ...591L..33L} and a mass of $2\times 10^7\,M_{\odot}$ \citep{2006ApJ...640..270S} is derived. Beside such large complexes which cover around 40\% of the sky for a resolution limit of $N_{HI}\ge 7\times 10^{17}\, \mbox{cm}^{-2}$ \citep{1995ApJ...447..642M}, there have also been found isolated, compact gas cloudlets with a small angular extent \citep{1999A&A...341..437B}. The first all-sky study of this sub-population, called Compact High Velocity Clouds (CHVCs), was presented by \citet{2002A&A...392..417D} as a combination of data extracted by \citet{2002A&A...391..159D} from the Leiden Dwingeloo Survey for the northern hemisphere and from the HI Parkes All Sky Survey (HIPASS), extracted by \citet{2002AJ....123..873P} for the southern hemisphere. Due to the compactness of CHVCs, the lack of distance indicators, and the poor knowledge about their physical environment, direct distance measurements are difficult. 

Observations of M31 have shown recently that HVC-analogs are clustered around their host galaxy with a maximum distance between 50 kpc \citep{2004ApJ...601L..39T} and 60 kpc \citep{2005A&A...432..937W}. Most recently, \citet{2011A&A...528A..12R} have studied Ca II absorbers at low redshift and propose that the characteristic radial extent of HVC analogs with $\log N(HI)\ge 17.4$ in galaxies with $z<0.5$ is $R_{HVC}\approx 55\,\mbox{kpc}$. 

Another remarkable property of HVCs is a 2-phase structure with a cold, dense core and a warmer, more tenuous envelope, which was already discussed by \citet{1991A&A...250..484W} for clumps in the larger complexes and is still supported by more recent observations of smaller cloudlets as in \citet{2006A&A...457..917B}. \citet{2010ApJ...708L..22G} presented ALFALFA (Arecibo Legacy Fast ALFA) observations of ultra-compact high velocity clouds (UCHVCs) at assumed extragalactic distances. The derived masses of the measures UCHVCs at a distance of 1 Mpc are below $10^6\, \Msun$, which explains the non-detection by previous HI surveys, such as \citet{2007ApJ...662..959P}. 

\subsubsection{Dark Matter Minihalo and Ram Pressure Stripping: \\}

If the CHVCs permeate the intergalactic space their origin allows two possibilities:  they are disrupted remnants from galaxy collisions and formed in the tidal tails or they represent the relics of structure formation in a CDM cosmology. The latter possibility would provide the requested link to the postulated DM building blocks of hierarchical galaxy formation which means that this baryonic cloud mass should be harbored by a so-called DM mini-halo. \citet{2009ApJ...707.1642N} have investigated the orbit of the Smith Cloud under the consideration that the cloud is confined by a DM halo.

Detailed observations of HVCs are showing a head-tail structure in the hydrogen column density distribution in both over-densities within the large complexes \citep{2000A&A...357..120B} and single CHVCs, like presented for example by \citet{2001A&A...370L..26B} and \citet{2006A&A...457..917B}. In agreement with theoretical and numerical studies this structural properties are unambiguously indicative for gas clouds moving through a surrounding plasma.
It is expected that if the force due to ram pressure exceeds the gravitational restoring force that keeps the cloud material via self-gravity together, material is stripped off the cloud \citep{1972ApJ...176....1G}. This leads to a head-tail structure of the cloud, which is observed with radio observations of the 21cm line of neutral hydrogen \citep{2000A&A...357..120B,2002AJ....123..873P,2005ASPC..331..105W}. Furthermore, the mixing of hot halo and cool cloudy gas might be responsible for the observed ionization stages in the hot halo \citep{1996ApJ...458L..29S}.

Observationally derived structure parameters allow the conclusion that CHVCs are bound by a DM halo of significant mass \citep{2000A&A...354..853B}. While the gravitational potential in the vicinity of the HVC is dominated by the combined gravitational potentials of the cloud itself and the host galaxy, a DM halo around the CHVC enlarges the total mass of the cloud-halo system and therefore also the volume in which the cloud potential exceeds the potential of the major galaxy, so that HVC material is gravitationally stronger bound to the cloud. 

\begin{table*}[htdp]
\caption{Initial parameters of the high-velocity clouds in different simulations. First column: initial setup no. for every run; col. 2: relative HVC velocity $v_w$; col. 3-5: HVC temperature $T_{gas}$, radius $R_{gas}$ in pc, and mass $M_{gas}$ in $\Msun$; 
DM subhalo scale radius of $r_0$ (col. 6) in pc and mass $M_{DM}$ (col. 7) in $\Msun$; col. 8: baryon-to-total mass ratio $M_{gas}/M_{tot}$;  maximum refinement level of the adaptive grid (Max.Ref.) in col. 9 and effective resolution (Res.) in col. 10. In the last column (Run Nr.) a unique number is assigned to every simulation run.}

\begin{center}
\begin{tabular}{lcccccccccr}
\hline
\hline
Setup & $v_{w}$ & $T_{gas}$ & $R_{gas}$	& $M_{gas}$ & $ r_0$ & $M_{DM}$	 & $M_{gas}/M_{tot}$ & Max.Ref. & Res. & Run\\

	& [$km\,s^{-1}$] & $[K] $  & $[pc]$ & $[\Msun]$	& $[pc]$ & $ [\Msun]$ & {\bf ($M_{tot}=M_{gas}+M_{DM}$) }  &  & [pc] & Nr.\\

\hline
		
2	& 200&	1000	&	194	&	$8.55 \times 10^5$	&	   -	&	-				&	-	& 5	& 38	&I \\	
	& 200&	1000	&	194	&	$8.55 \times 10^5$	&	150	&	$4.93\times 10^6$	&	0.15	& 5	& 38	&II\\	
\hline
2-D	& 200&	1000	&	125	&	$4.12 \times 10^5$	&	-	& 	-				&	-	& 5 	& 38	&III \\	
	& 200&	1000	&	125	&	$4.12 \times 10^5$	&	150	&	$4.93\times 10^6$	&	0.08	& 5	& 38	&IV\\	
\hline
2-DD& 200&	1000	&	127	&	$4.40 \times 10^5$	&	-	&	-				&	- 	& 6	& 25	&V\\	
	& 200&	1000	&	127	&	$4.40 \times 10^5$	&	-	&	-				&	- 	& 7	& 5	&VI\\	
	& 200&	1000	&	127	&	$4.40 \times 10^5$	&	150	&	$4.93\times 10^6$	&	0.08 	& 7	& 5	&VII\\ 	
	& 200&	1000	&	127	&	$4.40 \times 10^5$	&	250	&	$1.63\times 10^7$	&	0.03 	& 6	& 25	&VIII\\	
	& 200&	1000	&	127	&	$4.40 \times 10^5$	&	250	&	$1.63\times 10^7$	&	0.03 	& 7	& 5	&IX\\	
\hline
	& 250&	1000	&	127	&	$4.40 \times 10^5$	&	-	&	-				&	- 	& 6	& 18	&X\\	
	& 250&	1000	&	127	&	$4.40 \times 10^5$	&	250	&	$1.63\times 10^7$	&	0.03 	& 6	& 18	&XI\\	
\hline
3	& 200&	2000	&	376	&	$3.10 \times 10^6$	&	-	&	-				&	- 	& 4 	& 101&XII\\  	
	& 200&	2000	&	376	&	$3.10 \times 10^6$	&	-	&	-				&	- 	& 5 	& 50	&XIII\\ 	
	& 200&	2000	&	376	&	$3.10 \times 10^6$	&	-	&	-				&	- 	& 6 	& 25	&XIV\\ 	
	& 200&	2000	&	376	&	$3.10 \times 10^6$	&	-	&	-				&	- 	& 7 	& 13	&XV\\ 	
	& 200&	2000	&	376	&	$3.10 \times 10^6$	&	150	&	$4.93\times 10^6$	&	0.39 	& 5 	& 50	&XVI\\  	
	& 200&	2000	&	376	&	$3.10 \times 10^6$	&	150	&	$4.93\times 10^6$	&	0.39 	& 6 	& 25	&XVII\\  	
	& 250&	2000	&	376	&	$3.10 \times 10^6$	&	-	&	-				&	- 	& 5 	& 50	&XVIII\\ 	
	& 250&	2000	&	376	&	$3.10 \times 10^6$	&	150	&	$4.93\times 10^6$	&	0.39 	& 5 	& 50	&XIX\\ 	
\hline
3-D	& 200&	2000	&	230	&	$1.39 \times 10^6$	&	-	&	-				&	- 	& 5	& 50	&XX\\ 	
	& 200&	2000	&	230	&	$1.39 \times 10^6$	&	150	&	$4.93\times 10^6$	&	0.22 	& 5	& 50	&XXI\\ 	
	& 200&	2000	&	230	&	$1.39 \times 10^6$	&	150	&	$4.93\times 10^6$	&	0.22 	& 6	& 25	&XXII\\ 	
\hline
\hline
\end{tabular}
\end{center}
\label{tab:overview}
\end{table*}%

In addition, whether HVCs represent the baryonic content of cosmological DM sub-halos or not, should become visible during the ram pressure stripping (RPS) action on their passage through the hot halo gas. The gravitational acceleration in the vicinity of the HVC is dominated by the self-gravity of the cloud and the external gravitational potential of its host galaxy. At first, the cloud gas is confined within the equipotential of cloud and galaxy potential. Once the stripped-off material runs over this region, it gains an additional acceleration in comparison to the material that is stripped off but still gravitationally bound to the cloud. This effect is expected to be hardly noticeable if the HVC's total mass is only determined by its gas mass, because the potential of the massive host galaxy mass would widely dominate the self-gravity of the cloud. If the cloud is embedded in a DM subhalo the total mass of the HVC and therefore its self-gravity is much higher and could make this effect visible in velocity-position space as a sharp change in the velocity gradient along the head-tail structure of the cloud, as observations obtrude \citep{2000A&A...357..120B,2001A&A...370L..26B,2006A&A...457..917B}.

\subsection{Numerical Simulations}

\subsubsection{Ram Pressure Stripping: \\}

\citet{2001ApJ...555L..95Q} have studied by means of hydrodynamical simulations the behavior of CHVCs passing through the dilute hot halo gas of a typical disk galaxy at large distances. The main purpose of their numerical models was to compare the emerging head-tail structure with observations, in order to draw conclusions on the necessity of a DM component and on a possible lower density limit for the surrounding intergalactic medium (IGM). 

They found head-tail structures in environments with a density higher than $10^{-4}\, cm^{-3}$, whose lifetimes in simulations without a DM halo are very short, $\approx 10\,$Myrs only, while the tails in their DM dominated clouds survive for $\approx 1\,$Gyr.

Although the authors have simulated a grid of HVC models, it is worth to point out, that there is no comparison of the same cloud with and without DM, like we discuss in this work, but the authors study two different scenarios: HVCs as close-by structures at a distance of 1-10 kpc and masses between 10 and 100 $M_{\odot}$ vs. very distant DM dominated clouds at a distance of 300 kpc, a radius of 1 kpc and a mass of more than $10^7\, M_{\odot}$, contrary to the typical maximum distance of around 55 kpc \citep{2011A&A...528A..12R}.

A snapshot of the DM dominated cloud modeled over 300 Myrs (Fig. 2 of \citet{2001ApJ...555L..95Q}) with a constant velocity of 200 km/s travels over a distance of 60 kpc through the galactic halo, i.e. must have faced changes in the ambient densities, a tidal field and an acceleration by the host galaxy. Due to the lack of self-gravity, low-mass DM-free gas clouds are disrupted very quickly so that the evolution is only shown after 3 Myrs \citep[][Fig. 3]{2001ApJ...555L..95Q}. The DM-free clouds in the simulation runs presented in this work can easily survive for 100 Myrs, not only because of the included self-gravity but also due to the much higher masses and distances, in agreement with recent observations. 

\citet{2009ApJ...698.1485H} have shown with grid-based hydrodynamic simulations that smaller clouds with HI masses $<10^{4.5}\, \Msun$ will lose their HI content during their paths of 10 kpc at maximum (for typical relative velocities and halo densities). They propose that the stripped material which is usually gas in the warm and ionized phase  can contribute to the extended layer of warm, ionized gas at lower galactic heights \citep{1993AIPC..278..156R, 2008PASA...25..184G} and later on recool and form smaller HI cloudlets which are classified as low or intermediate velocity clouds (LVC, IVCs).

\subsubsection{Heat Conduction: \\}

Due to the large temperature gradient between the cloud and the surrounding halo gas, saturated heat conduction should be an important process of energy transport between the gas phases, leading to the classical two-phase structure. Furthermore, heat conduction stabilizes clouds with head-tail structure and is extending their lifetime \citep{2007A&A...472..141V}. In addition, \citet{2007A&A...475..251V} have shown that in models which combine heating, cooling and saturated heat conduction with self-gravity, condensation of halo gas on the cloud surface can dominate evaporation and lead to a net accretion of ambient gas. 

For the warm cosmic rain theory by \citet{2009ApJ...698.1485H} this means that the stripped off material is not only warmer and higher ionized than the cold core of HVCs, but would have other significant consequences: On average HVCs contain much smaller element abundances than the Milky Way ISM (see Tab.~\ref{tab:metall}) what clearly advocates for their primordial origin. HVCs affected by heat conduction and stripping in the presumably metal-rich halo gas are expected to have also a higher metallicity because the stripped cloud surface consists not only of cloud gas but also of accreted halo gas. If part of the LVC and IVCs are the decelerated survivors of disrupted or ablated HVCs they should have a higher metallicity than the less processed HVCs, which is observationally confirmed (Tab.~\ref{tab:metall}). Since IVCs/LVCs rain down with velocities below escape velocity, an origin from the galactic fountain cannot be ruled out. The cores of HVCs, which survive ram-pressure stripping, can be decelerated by either the drag of the hot halo gas or other processes like the buffering of plane-parallel galactic magnetic field lines which have to couple to the HVC plasma \citep{1997A&A...320..746Z}. This effect would contribute to the heating of the halo gas \citep{2011CoPhC.182.1784J}.

In a recent study about the origin of the HVCs \citet{2009MNRAS.397.1804B} suggested that they can be formed by thermal instability in the warm-hot Galactic halo medium. For Milky Way sized halos, however, they found that this happens only in regions outside of 100 kpc and for an almost perfectly flat entropy profile, when thermal instabilities are not damped by a combination of buoyancy and thermal conduction. They concluded that it is therefore a rather unlikely formation scenario for HVCs.

\begin{table}[tdp]
\caption{ The initial distribution of the cloud follows in all setups a hydrostatic equilibrium, which is derived by Eq.~\ref{equ:hydro2}, where $M(r)$ includes either only the gas mass or in addition to that also the enclosed DM mass following Eq.~\ref{equ:burkert1} and \ref{equ:burkert2} with a scale radius of $r_0 = 150\, pc$ or $r_0 = 250\, pc$. } 
\begin{center}
\begin{tabular}{lll}
\hline
\hline
Setup 		&		T [K] 	&	$M(r)$ \\
\hline
2			&		1000 K		&	gas only		\\
2-D			&		1000 K		&	gas + DM halo ($r_0 = 150\,pc$)	\\
2-DD		&		1000 K		&	gas + DM halo ($r_0 = 250\,pc$)	\\
3			&		2000 K		&	gas only		\\
3-D			&		2000 K		&	gas + DM halo ($r_0 = 150\,pc$)	\\
\hline
\end{tabular}

\end{center}
\label{tab:setups}
\end{table}%

\subsection{General Questions} 

Since differences in the ram pressures stripped gas and in the head-tail structure are not yet explored with respect to their possible origins, here we present numerical models of HVCs with and without a stabilizing DM subhalo based on the following questions:
Is it possible for the cloud with DM to keep the gas longer bound as without? Does a DM halo suppress hydrodynamical instabilities? Is a cloud without a DM halo already disrupted and does it lose most of its gas due to the ram pressure produced by its supersonic motion (up to more than $400\kms$) through the hot gaseous halo of its host galaxy or in the outskirts of the dilute halo gas? Can we distinguish from the observed velocity - position distribution of the head-tail structure whether the observed cloud contains DM? If they are embedded in a DM halo, are HVCs the low mass extensions of gas rich subhalos with (ultra faint) dwarf galaxies as their higher mass analogs? Moreover, if they are DM-free and a significant fraction of the few observed satellite galaxies have a tidal origin \citep{2006A&A...456..481B,2000ApJ...543..149O}: where are all the DM satellites we see in large cosmological simulations?

\begin{figure}[htbp]
\begin{center}
		 \includegraphics[width=\linewidth]{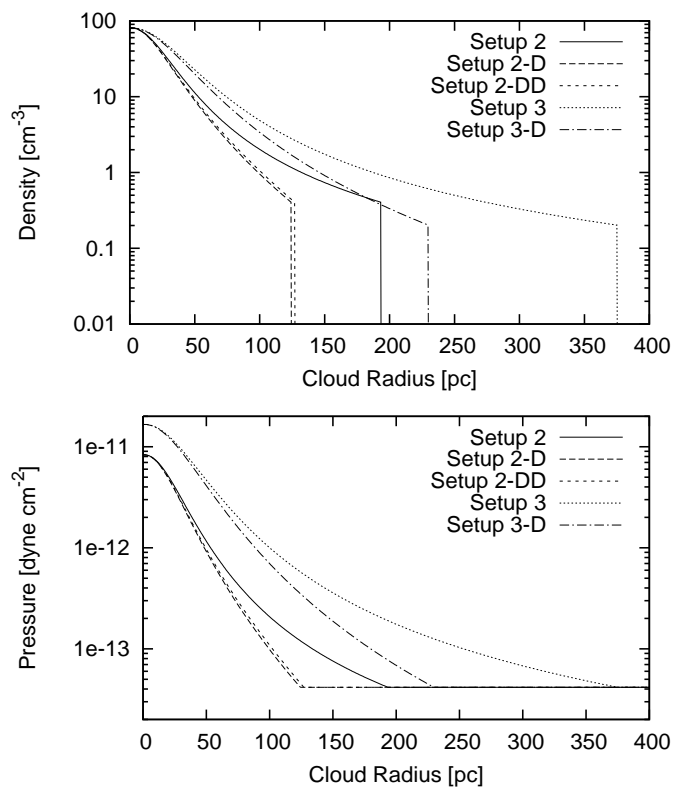}
\caption{Initial density (top panel) and pressure (lower panel) distribution of the model setups described in Tab.~\ref{tab:overview}.}
\label{fig:overview}
\end{center}
\end{figure}

\subsection{Paper Structure}

This paper has the following structuring:
First the numerical method and the model setup of both the host galaxy and the HVCs are presented. In Sect.~\ref{sec:lagrange} we discuss the size of the Roche-lobes in this 2-body system and the consequences for the gas that is lost by RPS. 

The effect of RPS that leads to the observed cometary appearance of the clouds is discussed in Sect.~\ref{sec:rampres}, while further possible mass-loss procedures are mentioned in Sect.~\ref{sec:other}.

Because HVCs do not necessarily fall directly towards the observer nor radially into the center of the Milky Way we take a close look on the line-of-sight problem in Sect.~\ref{sec:lineofsight}. Not only the column density distribution varies when changing the inclination between object and observer but the velocity measurements can cover only the radial velocity. Therefore the whole phase-space distribution of the observed clouds is highly sensitive to line-of-sight effects. For all performed simulations we processed the data in a way that we simulate different line of sight directions to look for signatures that are dependent on the line of sight. 

Finally, the dependency of the results on parameters like the initial velocity (see Sect.~\ref{sec:windvel}) and the DM mass (see Sect.~\ref{sec:dmmass}) and shape (Sect.~\ref{sec:dmshape}) are discussed.

We have included not only a self-gravitating gas cloud as well as a static gravitational potential that represents a DM mini-halo but also an external gravitational potential which gives us the opportunity to study the equipotential lines of the combined potential of the galaxy and the HVC. 

\begin{figure*}[tbp]
 	\begin {minipage}[b]{0.47\linewidth}
		\includegraphics[width=\linewidth, bb = 0 0 566 255,clip]{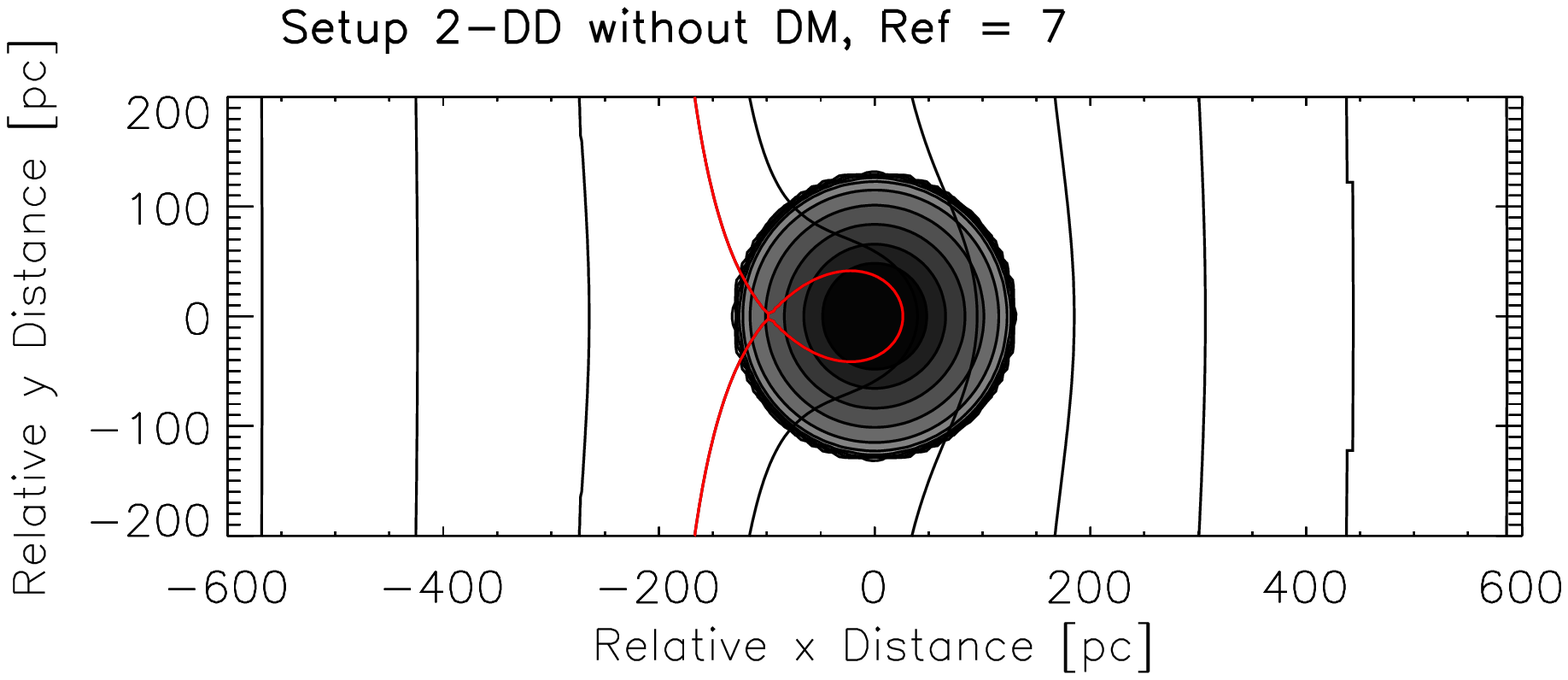}
	\end {minipage}
	 \begin {minipage}[b]{0.47\linewidth}
		\includegraphics[width=\linewidth, bb = 0 0 566 255,clip]{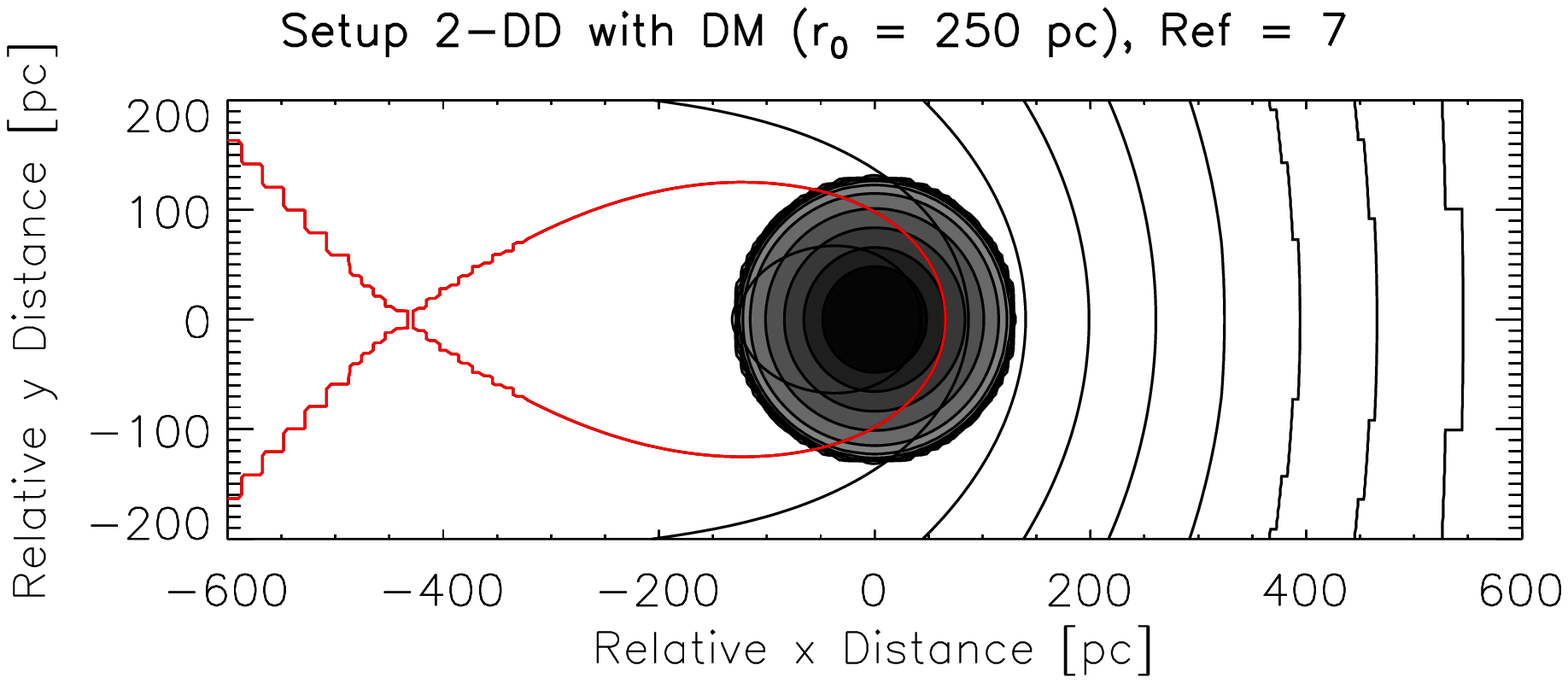}
	\end {minipage}

 	\begin {minipage}[b]{0.47\linewidth}
		\includegraphics[width=\linewidth, bb = 0 0 566 255,clip]{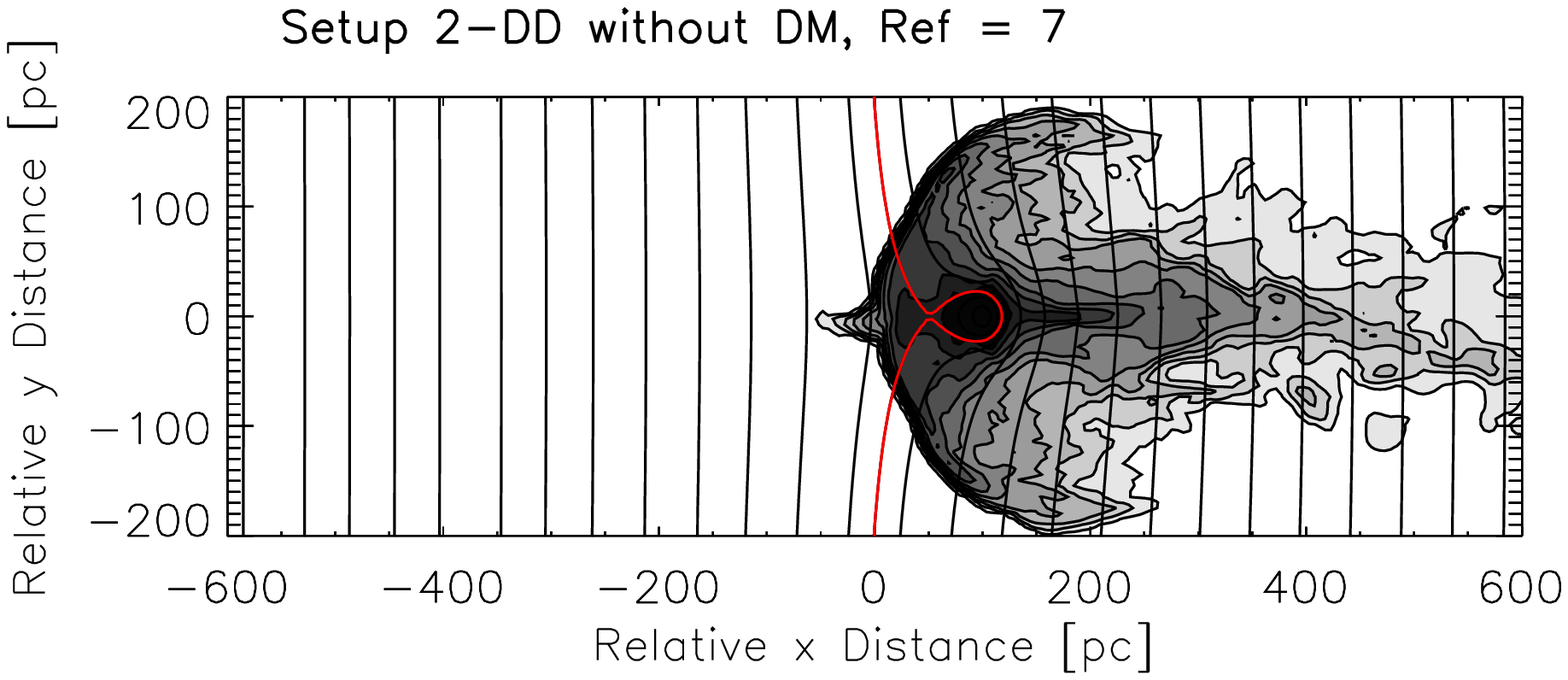}
	\end {minipage}
	 \begin {minipage}[b]{0.47\linewidth}
		\includegraphics[width=\linewidth, bb = 0 0 566 255,clip]{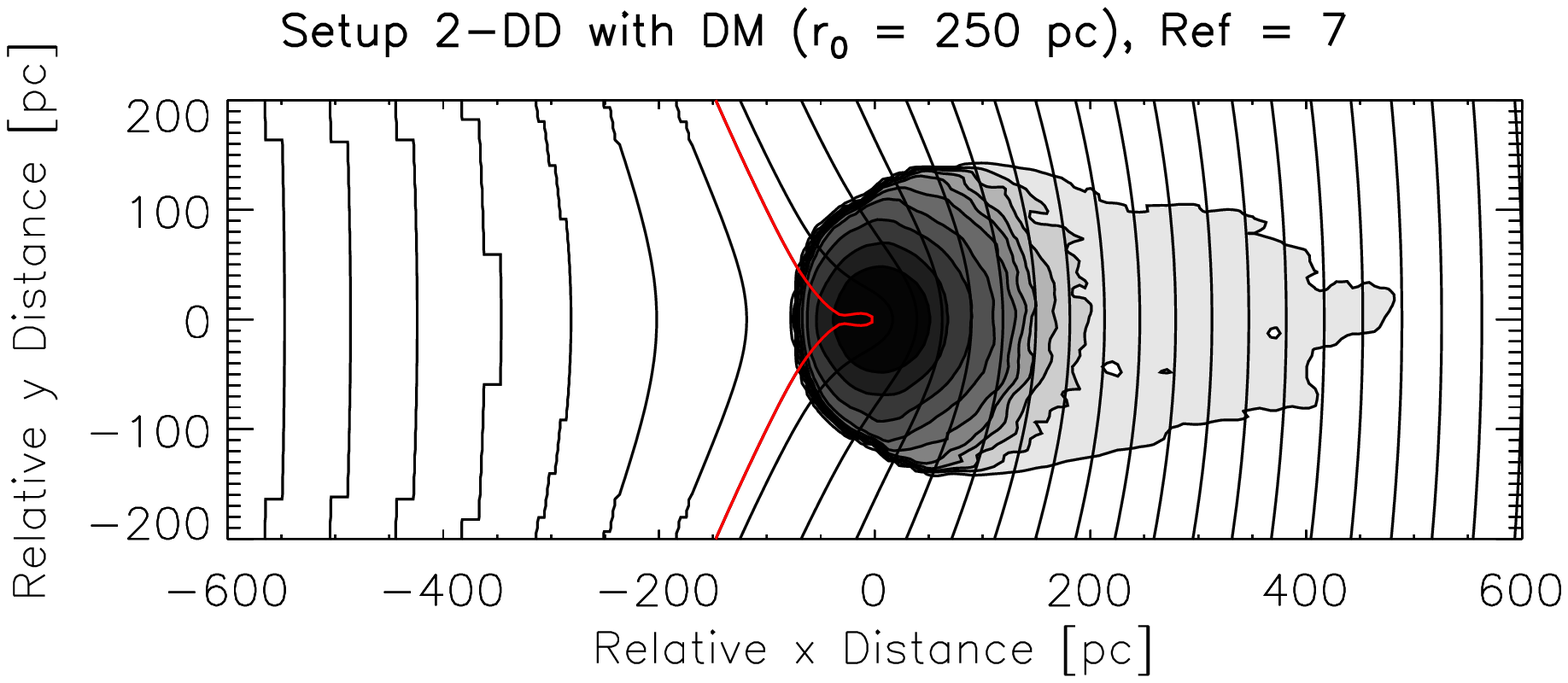}
	\end {minipage}
	\caption{Column density distribution of the gas cloud overlaid with equipotential lines for Setup 2-DD (see Tab.~\ref{tab:overview} for its parameters) at a maximum refinement level of 7 (effective spatial resolution: 5 pc): without DM halo (left panels, Run VI) and with DM halo of scale radius $r_0=250\,pc$  (right panels, Run IX). The contour lines of the greyscale \HI column density distribution range between $\mbox{log}(N_{\HI}) = 18.8 \mbox { ... } 21.2 \, cm^{-2}$ in steps of 0.3 dex. Overlaid are the equipotential lines of the total gravitational potential by the gaseous cloud, the DM subhalo of the cloud (if present), and the external galaxy (see text) in steps of $2.5\times 10^{11}\,cm^{2}\,g^{-2}$. In red, the equipotential line of the innermost Lagrangian point is depicted. Top panels show the initial distributions, bottom panels after a simulation time of 100 Myrs. (Step-like contour lines reflect the coarser grid further away from the cloud.)}
\label{fig:rochelobe2dd}
\end{figure*}

The initial distance between the HVC and the external mass was chosen to be 50 kpc, which leads to cloud masses in the order of $10^5$ to $10^6$ solar masses. Outside of 60 kpc the HVCs are either fully ionized and therefore not visible in neutral hydrogen or HVCs origin from an interaction scenario like the Magellanic Stream which would also explain the typical maximum distance of around 50 - 60 kpc.

\section{\label{sec:models}Models}

\subsection{\label{sec:setup}Numerical Methods}

In order to resolve hydrodynamical instabilities and turbulence in the stripped-off gas, but not to waste computational time and spatial resolution by the dynamics outside of the clouds, an adaptive-mesh algorithm is a proper choice.  

For all simulations in this project the hydrodynamic code FLASH3.2 developed by \citet{2000ApJS..131..273F} is used, which already provides a fast multigrid Poisson solver which is based on \citet{2008ApJS..176..293R} and adapted for the FLASH grid structure by the Flash Center.

\subsection{\label{subsec:setup}Initial Setup}

The initial setup for every run is an isothermal cloud in hydrostatic equilibrium with temperature $T_{gas}$, radius $R_{gas}$ and mass $M_{gas}$ embedded in a \citet{1995ApJ...447L..25B} halo with a scale radius of $r_0$ and a mass  $M_{DM}$. The baryon-to-total mass ratio for the cloud is given by $M_{gas}/M_{tot}$ and the relative velocity between the cloud and the ambient medium is given as $v_{w}$. 

The simulation of an object moving towards the center of an external gravitational potential representing the host galaxy, is performed in a way, so that the rest frame remains at the cloud's mass center. This leads to a wind-tunnel setup with an increasing wind velocity, because the gravitational attraction of the host galaxy increases as the cloud approaches. Since observations do not show any star formation in HVCs, neither nowadays nor at any point in the past, the simulated cloud should not collapse due to self-gravity during the simulation time, but also not expand and dissolve at the simulated heights because we focus on the compact, isolated sub-class of the HVCs. Therefore we assume an initial hydrostatic equilibrium:

\begin{equation}  \label{equ:hydro1}
	\frac{\partial P}{\partial r} = -\rho(r) \frac {\partial \Phi(r)}{\partial r} 
\end{equation}

\noindent where $\Phi(r)$ is the gravitational potential of the cloud
and since the cloud is spherically symmetric in the beginning the pressure distribution follows: 

\begin{equation}  \label{equ:hydro2}
	\frac{\partial P}{\partial r} = -\rho(r) \frac {\partial }{\partial r} \frac {G M(r)}{r}
\end{equation}

\noindent where $\rho (r)$ is the gas density and $M(r)$ is the enclosed mass within the radius $r$. 

We intend to compare clouds that are embedded in a DM halo to clouds that only consist of baryonic matter. In these two cases the initial pressure and therefore density distribution are different since the enclosed mass contains both dark and baryonic matter. Subsequently other parameters as the cloud radius and the cloud mass differ between the cases with and without a DM subhalo. 

To avoid effects that are due to the different initial distributions we set up clouds that are in hydrostatic equilibrium with the gaseous part only (Setup 2 and Setup 3) as well as clouds that are stable under both the baryonic and the DM mass (Setup 2-D, Setup 2-DD and Setup 3-D) 

\begin{figure*}[t!]
		\rotatebox{90}{\includegraphics[height=0.45\linewidth, bb = 20 50 405 650,clip]{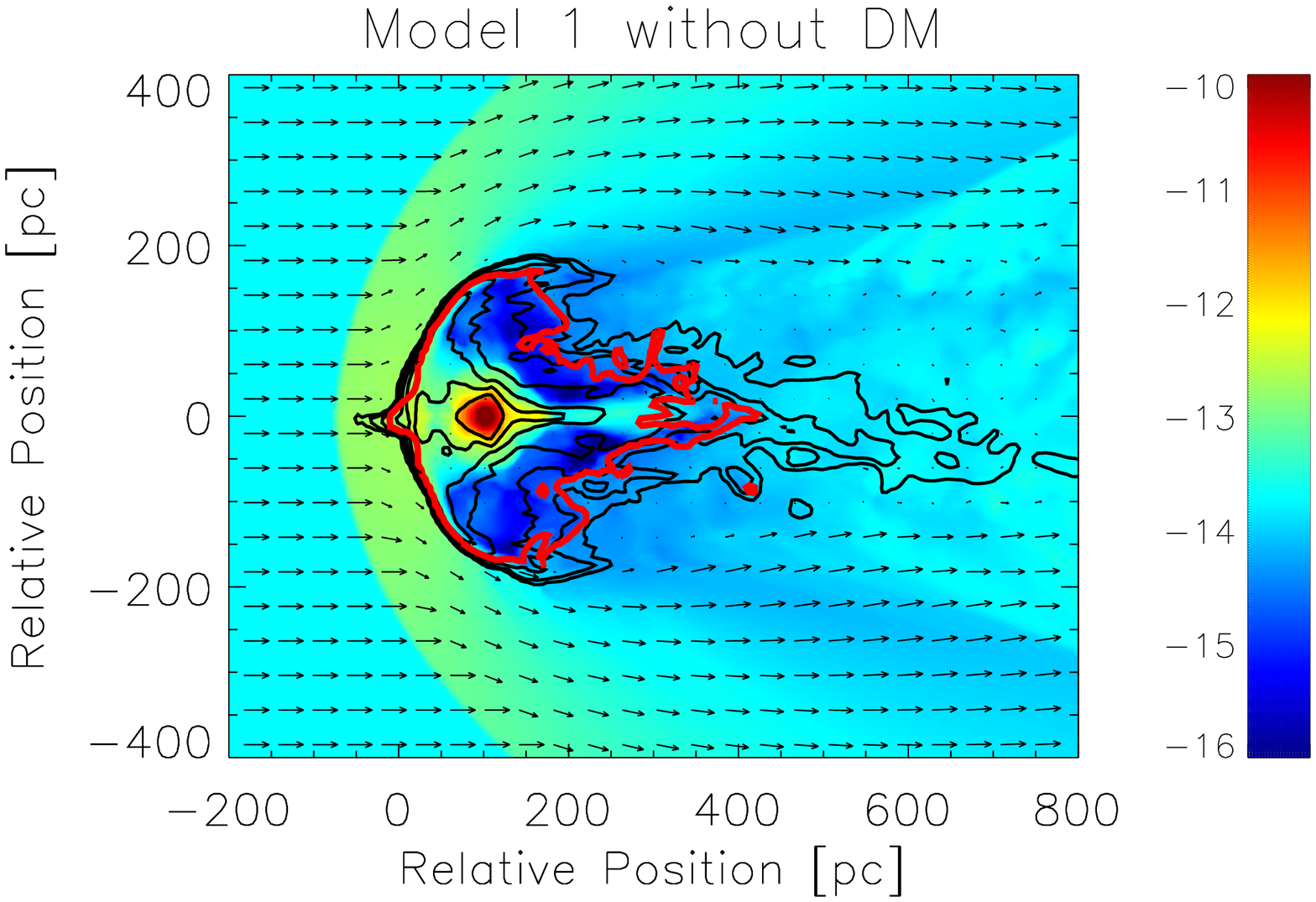}} 
		\rotatebox{90}{\includegraphics[height=0.45\linewidth, bb = 20 50 405 650,clip]{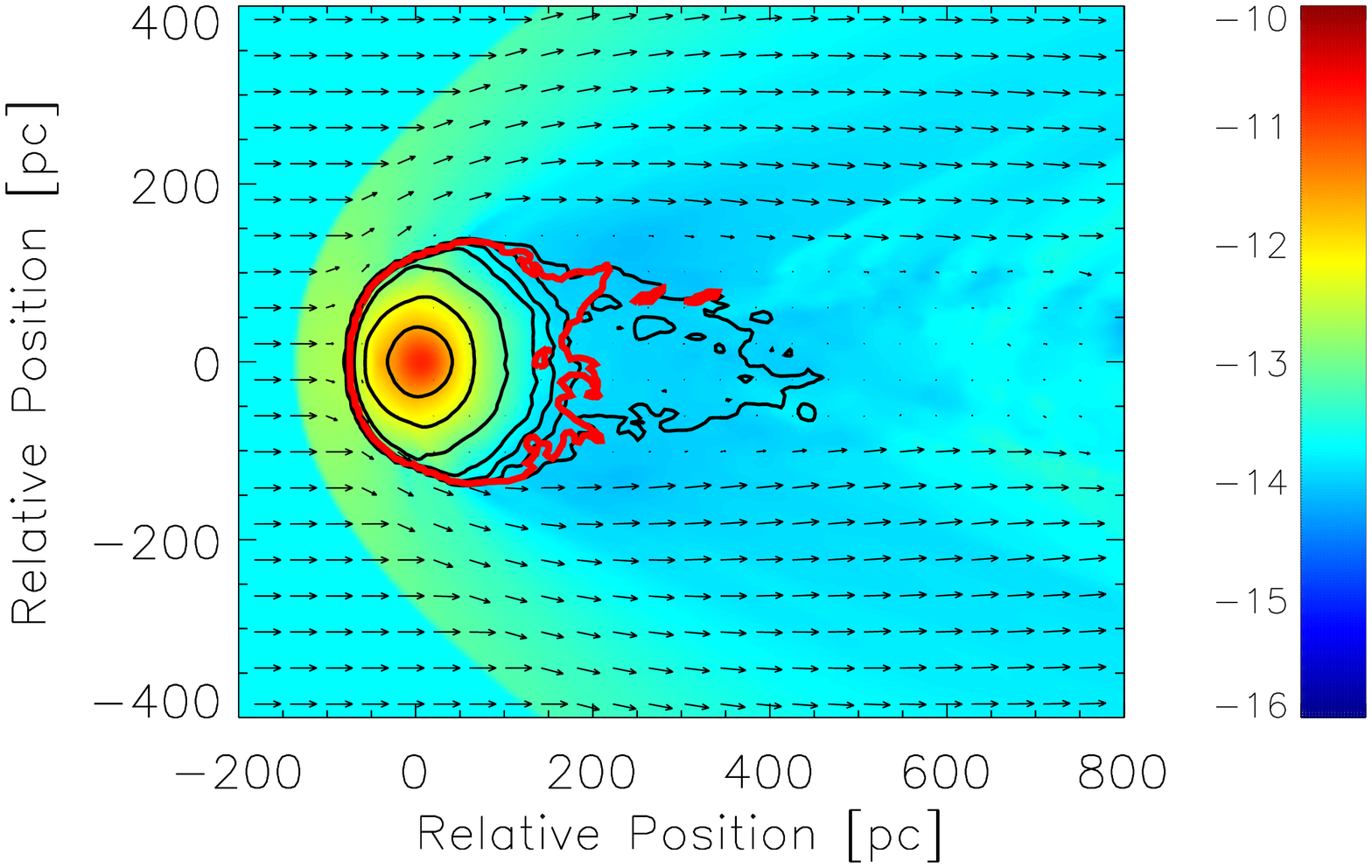}} \\
		\rotatebox{90}{\includegraphics[height=0.45\linewidth, bb = 20 50 405 650,clip]{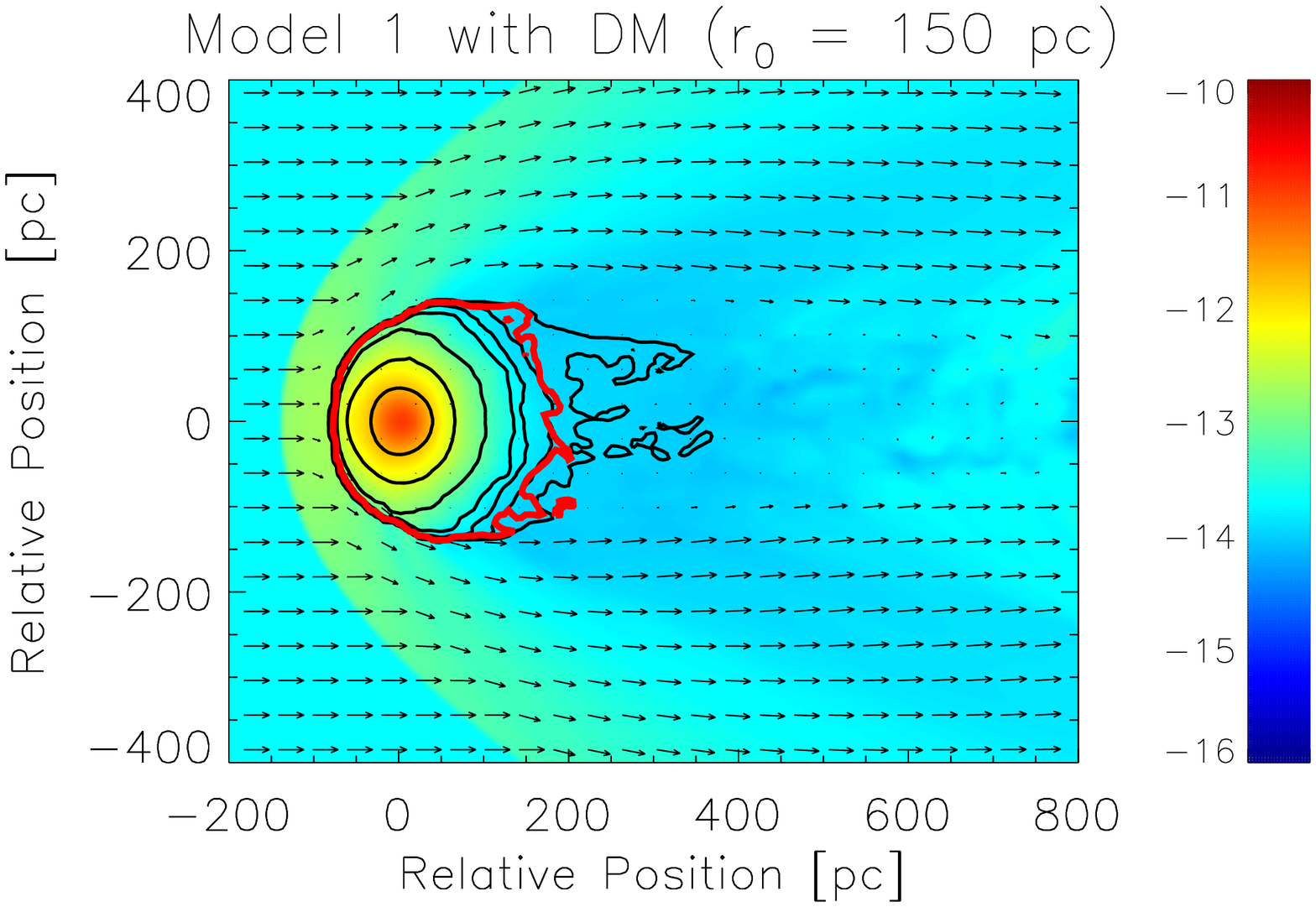}}
		\rotatebox{90}{\includegraphics[height=0.45\linewidth, bb = 20 50 405 650,clip]{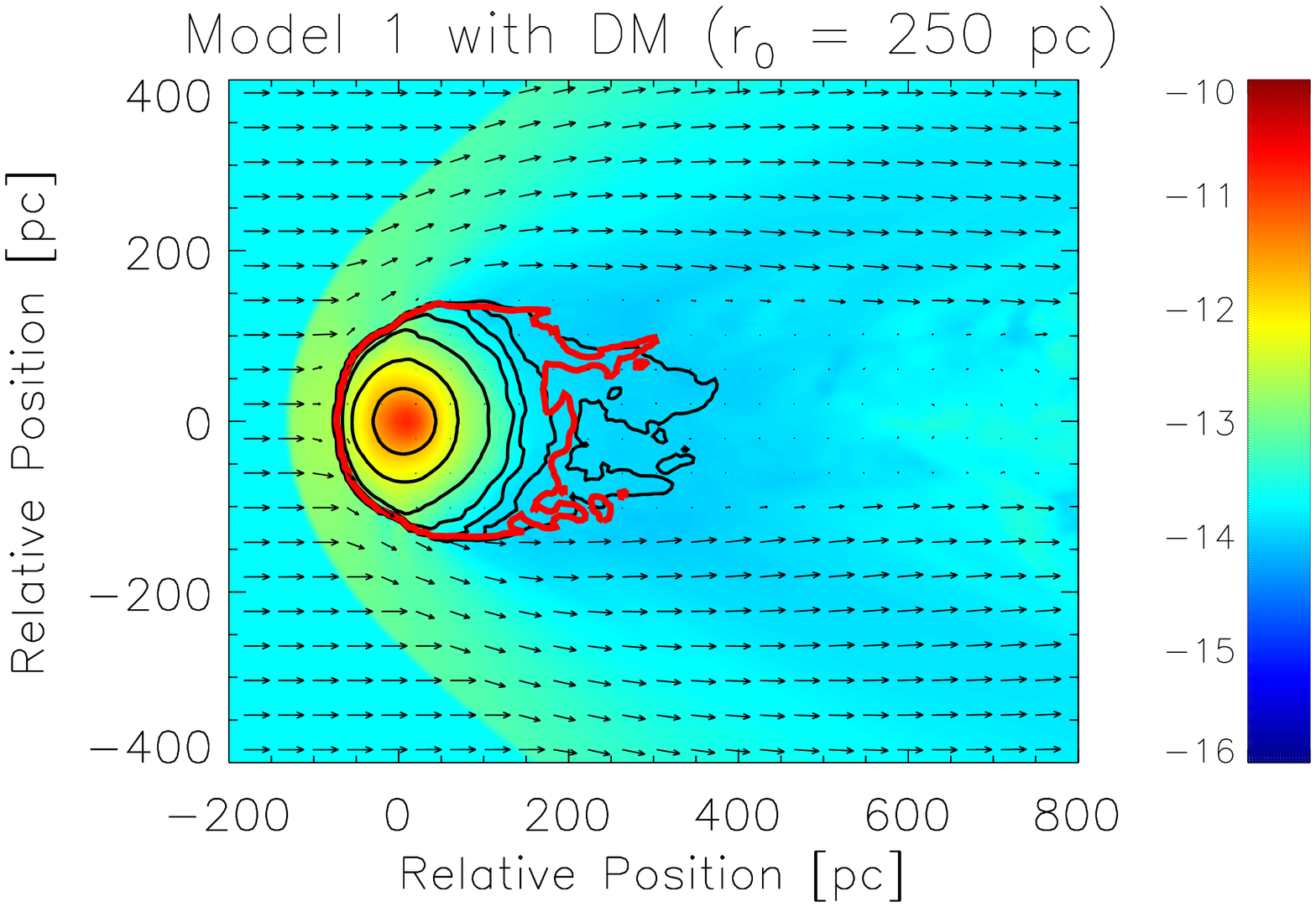}}
\caption{ HVC model with an initial gas mass of $M_{gas} = 4.4\times10^5\,\Msun$ (Setup 2-DD): color-coded pressure distribution (logarithmic scale  [dyne cm$^{-2}$]) through the midplane (z=0) of the 3D simulation box overlaid with \HI column density contours between $\mbox{log}(N_{HI}) = 18.5 \dots 21.5\, cm^{-2}$ in steps of 0.5 dex, 100 Myrs after the onset. 
Upper left panel: cloud without DM  (Run VI), upper right panel: cloud embedded in an NFW halo (see Sec.~\ref{sec:dmshape}), lower panels: clouds with Burkert halos with $r_0 = 150\,pc$ (Run VII, left panel) and $r_0 = 250\,pc$ (Run IX, right panel), respectively.
The red contour depicts the $(E_{kin}+E_i) = - E_{pot}$ boundary and encloses all material, both from the cloud and from the hot halo gas, that is gravitationally bound to the HVC. The velocity vectors indicate a maximum velocity of $215\, km\,s^{-1}$ and their lengths are linearly scaled. }
\label{fig:nature2dd}
\end{figure*}

\noindent
and performed simulations with and without a DM subhalo for each setup (see Tab.~\ref{tab:overview} and Fig.~\ref{fig:overview} for details).

Stability studies without wind are performed to ensure that the timescale of the changes due to the deviation from the hydrostatic equilibrium is negligible, i.e. much larger with respect to the ram pressure timescale. 

The cloud is embedded in a homogeneous medium with a temperature of $T_{HIM} = 10^6\,K$ and a density of $\rho_{HIM} = 4\times 10^{-4} \, H\,cm^{-3}$ which represents the hot galactic halo. While observationally determined values for the outer halo of the Milky Way are rare, \citet{2001ApJ...555L..95Q} have shown that head-tail features only appear if the external medium has a density larger than $10^{-4}\,cm^{-3}$ and a relative velocity of $200\kms$.

The cloud's pressure is integrated following Eq.~\ref{equ:hydro2} down to a radius where the pressure of the cloud equals the pressure of the ambient medium. 

Since our study is aiming at exploiting whether HVCs would represent the extension of DM-dominated baryon assemblies from dwarf galaxies to the lower masses it seems reasonable to choose also as DM density distribution the empirically derived profile by \citet{1995ApJ...447L..25B}:

\begin{equation}  \label{equ:burkert1}
\rho_{DM} = \frac{\rho_0 r_0^3}{(r+r_0)(r^2+r_0^2)}
\end{equation}

with the empirical correlation between $\rho_0$ and $r_0$ of 
\begin{equation}  \label{equ:burkert2}
\rho_0 = 4.5\times 10^{-2}\left ( \frac{r_0}{\mbox{kpc}}\right )^{-2/3} \Msun\,\mbox{pc}^{-3}
\end{equation}

In contrast to the mostly referred DM profile, the so-called NFW profile \citep{1997ApJ...490..493N}, which was initially derived from large-scale cosmological simulations, with the characteristic cuspy center \citep[see also ][]{1999ApJ...524L..19M, 2004MNRAS.349.1039N}, the observed rotation curves of dwarf galaxies show nearly solid body rotation, which is preferably fitted with a central core in the DM distribution \citep[e.~g.][] {1995ApJ...447L..25B, 2002A&amp;A...385..816D, 2005ApJ...634L.145G, 2005ApJ...634..227D}. The resulting DM halos in more recent high resolution cosmological simulations \citep{2010MNRAS.402...21N} support central cores that can also be derived from a global DM-radius relation from spirals to dwarf spheroidal galaxies \citep{2010ApJ...717L..87W}. For further discussions on this topic see Sect.~\ref{sec:dmshape}.

We decided to run all our simulations up to a given simulation time of 100 Myrs, so that we can neglect the change in the ambient pressure, while the effects of the time-dependent external potential are well described in our simulations. The distance of the CHVC on radial trajectory is in our case between 30 kpc and 50 kpc. We expect that for clouds closer than that, the simplification of a radial infall is not valid anymore and an eccentric orbit close to the disk would lead to tidal stretching of the clouds as observed in the large complexes closer to the Galactic disk.  

In order to cover a large range of parameter space but in a reasonable expense of computational time and to perceive clear differences between DM-dominated and pure gas clouds we use only two different scale radii ($r_0 = 150$ pc and $r_0 = 250$ pc) for the DM profile. In combination with a sample of baryonic masses we obtain setups with very different baryon-to-total~mass ratios $f_b=M_b/ M_{tot}$ both including very low as well as very high baryon contents. 

The larger DM halo with $r_0 = 250$ pc has a total DM mass which is more than 3 times higher than that of the smaller DM halo with $r_0 = 150$ pc. Due to the different scale radii and the core shape of the potential, the ratio of the enclosed mass of the central part does not represent the full mass ratio of the total halos. But within 400~pc, a typical extent of the head-tail structure, the larger halo has already 70\% more DM mass enclosed than the smaller one, leading also to a difference in the gradient of the DM potential of 70\%. If the simulations were very sensitive to the scale radius of the surrounding DM halo, we would expect a significant difference in the results. 

\citet{2006MNRAS.371..401H} studied structure formation by means of hydrodynamic simulations for different spatial resolutions and find that low-mass haloes below a critical total mass of $M_c(z=0)=6.5\times 10^9 h^{-1}\,M_{\odot}$ can be baryon poor with $f_b$ as low as 0.01. We cover the full range from baryon poor haloes over haloes with the universal baryon-to-total mass ratio derived by WMAP3 data \citep[$f_b \approx 0.17$,][]{2011ApJS..192...18K} to systems with $f_b$ as high as 0.39 (Setup 3) to investigate the influence of a DM subhalo in a RPS scenario. The latter $f_b$ anticipates that the outermost part of the DM subhalo is already distorted by the Galactic tidal force.

\subsection{\label{sec:para}Parameter variants}

\begin{figure}[htbp]
		\rotatebox{90}{\includegraphics[width=0.6\linewidth, bb = 20 50 405 650,clip]{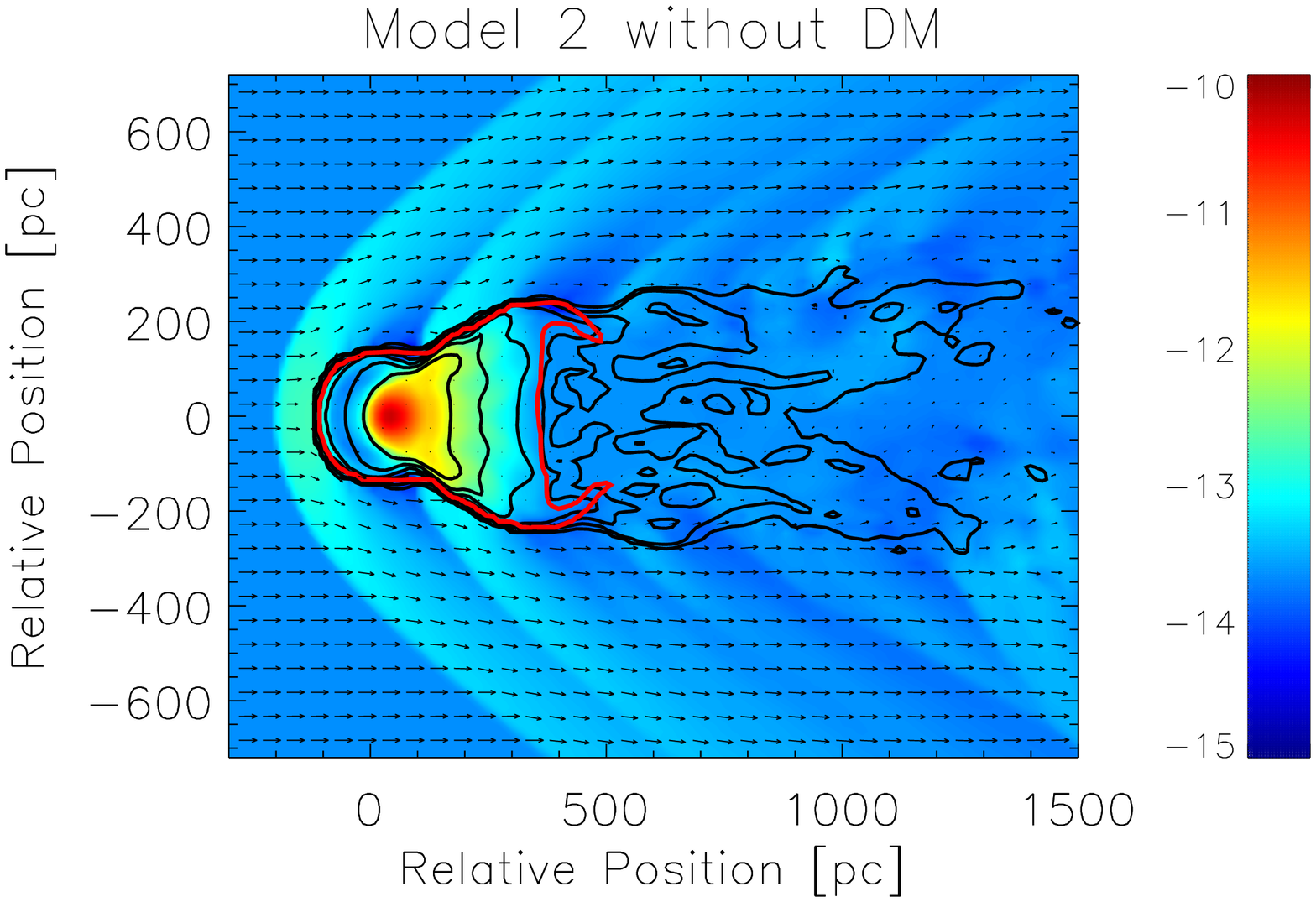}} \\
		\rotatebox{90}{\includegraphics[width=0.6\linewidth, bb = 20 50 405 650,clip]{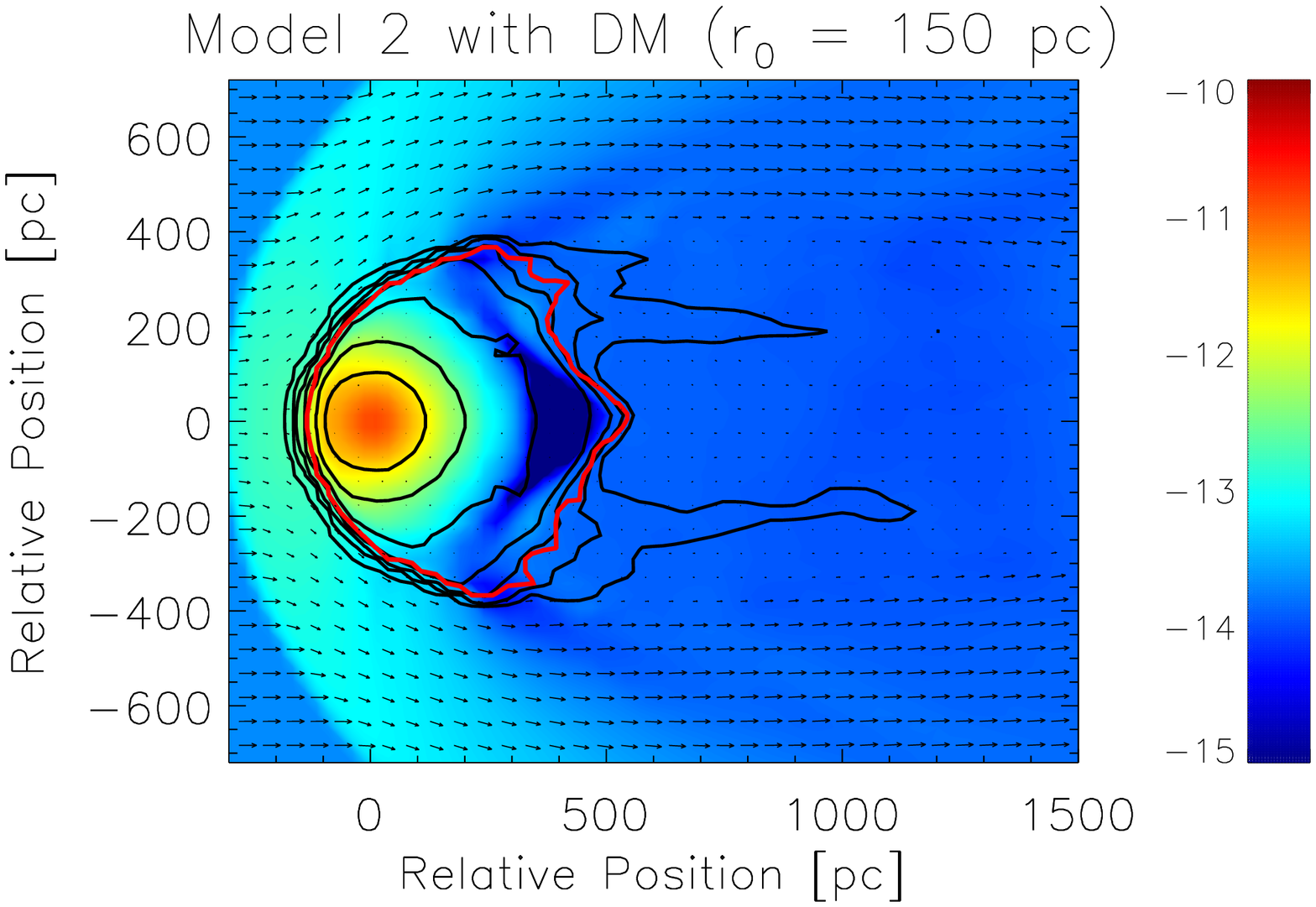}}

\caption{ As Fig.~\ref{fig:nature2dd}. The initial gas mass is $M_{gas} = 3.1\times10^6 \Msun$ (Setup 3). Upper panel:  DM-free cloud (Run XIV); lower panel: cloud with DM subhalo of 
$r_0 = 150 pc$ (Run XVII). }
\label{fig:nature3}
\end{figure}

For a better comparison between the simulation runs with and without DM content for every mass range runs with different stability setups are performed. One setup represents an initial distribution in hydrostatic equilibrium for a self-gravitating gas cloud (setup 2 and 3), while another setup is stable in a static run that also includes DM with a mass of $4.93\times 10^6 \Msun$  and a scale radius of 150 $pc$ for setups 2-D and 3-D and with a DM mass of $1.63\times 10^7 \Msun$ and a scale radius of 250 $pc$ for setup 2-DD (see Tab.~\ref{tab:setups}). For all these initial distributions we performed runs with and without a DM halo. This leads to configurations that are not in perfect hydrostatic equilibrium (see Sec.~\ref{sec:setup} for details). In Table~\ref{tab:overview} a summary of the used parameters can be found.

\subsection{\label{sec:lagrange}External gravitational potential}

Observations of individual CHVCs by \citet{2001A&A...370L..26B} and \citet{2006A&A...457..917B} show not only a head-tail structure in the column density distribution but also a clear change in the velocity gradient along the elongated density contours. 

Since the gravitational potential in the vicinity of the HVCs is dominated by its own mass and the external gravitational field of its host galaxy, this system can be considered as a two body interaction. Therefore the area within the equipotential line through the Lagrangian point 1 (in binary systems called ``Roche lobe") determines the region in which the cloud material is gravitationally bound to the HVC. Ram pressure leads to an elongation of the gas density profile in wind direction and as soon as gas has crossed the Roche lobe it becomes unbound from the HVC and should therefore be exposed to an additional acceleration visible in the immediate change of the velocity gradient. Since the size of the Roche lobe is dependent on the mass of the HVC, the observed position of the velocity gradient change can help to distinguish whether HVCs contain a significant fraction of DM or not. 

In our simulations the cloud will start at an initial distance of 50 kpc to an external point mass of $M_{Gal} = 10^{11}\, \Msun$ which represents the host galaxy. \citet{2008ApJ...684.1143X} have derived the Milky Way DM halo mass from the kinematics of blue horizontal branch stars in the galactic halo. They fitted both an unaltered NFW profile as well as an adiabatically contracted NFW halo and find virial masses between $M_{vir} = [0.82^{+0.21}_{-0.18}\, ...\, 1.21^{+0.40}_{-0.30}] \times 10^{12} \Msun$ with virial radii in the range of $r_{vir} = [258^{+20}_{-21}\,...\,293^{+31}_{-26}]\,kpc$. We compared the gravitational acceleration in the cloud region of the external point mass with $M_{Gal} = 10^{11}\, \Msun$ to the NFW profile represented by [$M_{vir} = 0.82\times 10^{12}\,\Msun$, $r_{vir} = 258\, kpc$, $c = 12.2$] and found that the relative difference is as small as 1-3 \% which is much less than the difference between the models by \citet{2008ApJ...684.1143X}. 

The equipotential lines together with the column density distribution of the isothermal gas distribution initially as well as after 100 Myrs of simulation time are shown in Fig.~\ref{fig:rochelobe2dd} for Setup 2-DD with (Run VIII and IX) and without DM (Run V).

In all setups the gaseous parts of the clouds exceed the smallest Lagrangian equipotential surface in at least one direction. In the setups without the additional mass of a DM halo, the L1 surface lies completely within the cloud. 
Note that the Roche lobes indicated in Fig.~\ref{fig:rochelobe2dd} become smaller during time not because of the mass loss but because the cloud approaches the galaxy mass with an initial velocity of $200\kms$ and reaches a distance of around 30 kpc after the simulation time of 100 Myrs.

Including an external mass in our simulations does not only effect the region where the cloud material is bound to the objects, but also influences the motion of the HVC. Due to the gravitation by a host galaxy the cloud is accelerated towards the external point mass. \citet{1997ApJ...481..764B} have shown that clouds reach a terminal velocity on their way to the galactic disk, where the acceleration due to gravitational force is equal to the drag force caused by the different densities of the hydrostatic layers of the galactic disk. Their model is valid for distances up to 10 kpc and they did not take the change in the area which is exposed to drag into account. In our setup the cloud travels in galactic heights between 30 kpc (after 100 Myrs of simulation time) and 50 kpc (initial distance), where the change of the hot halo gas density is negligible. The stronger gravitational acceleration at smaller distances and in many cases the shrinkage of the cloud surface that is exposed to drag due to RPS, together with a constant ambient medium does not lead to a terminal velocity in our models as assumed by \citet{1997ApJ...481..764B}.

\begin{figure*}[htbp]
\begin{center}
	\includegraphics[width=\linewidth]{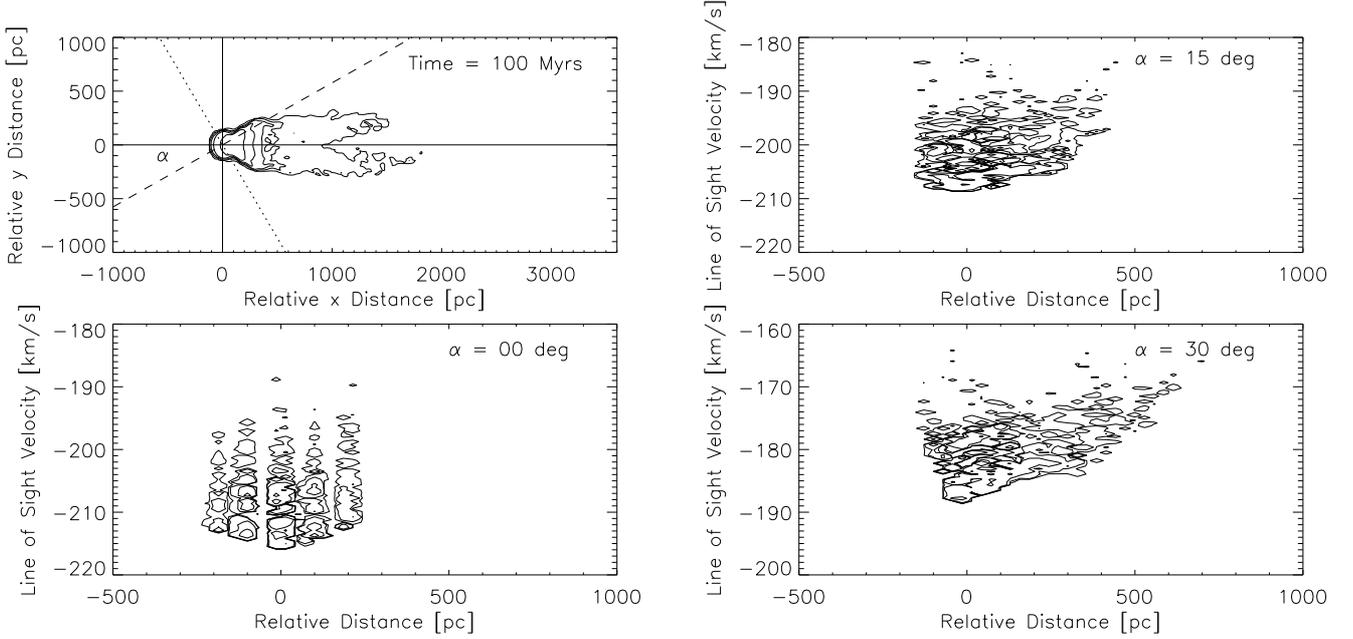}
\caption{ 
Projected velocity-space distribution of Setup 3 without a DM halo at a maximum refinement level of 7. {\it Top left panel}: contour lines of the column density distribution (in z direction) at $\log(N_{HI}) = 18.8\,...\,21.2\, H cm^{-2}$ steps of 0.6 dex 100 Myrs after the simulation onset. Both the x- and the y-coordinates correspond to the cartesian grid of the 3D data cube. Radial velocities measurable along the 
line-of-sight (dashed line) in different directions of angle $\alpha=0\,\deg, 15\,\deg$, and 30$\,\deg$ with respect to the x-direction are shown in the 3 frames from bottom left ($\alpha=0\deg$), through top right ($\alpha=15\deg$) to bottom right ($\alpha=30\deg$). Along the abscissa, projected distances perpendicular to the line-of-sight (dotted line) through the mass center (x=0) are shown. All cells with densities $n > 0.01\,H\,cm^{-3}$ are included and re-arranged to represent the column density distribution in the observed position-velocity space (leading to the bubbled contour structure). The contour lines of the line-of-sight column densities indicate $\log(N_{HI}) = 18,\,19,\,20 \mbox{ and }21\, H\,cm^{-2}$.
The measured maximum velocities differ accordingly from 215 (for $\alpha=0\deg$) to 188 $\kms$ ($\alpha=30\deg$). 
}
\label{fig:setup3_contourvel}
\end{center}
\end{figure*}

\subsection{\label{sec:rampres}Ram-pressure stripping}

We calculated the analytical stripping radius $R_{strip}$ as the radius where the force due to ram pressure $F_{ram}$ equals the gravitational restoring force $F_{grav}$. For radii larger than $R_{strip}$ the gravitational energy of the cloud mass is not large enough to keep the gas bound so that it is stripped off due to the relative motion of the hot ambient gas.
\citet{1972ApJ...176....1G} derived $R_{strip}$ for the instantaneous RPS of a gaseous disk moving face-on through the intra-cluster medium (ICM). \citet{2008MNRAS.383..593M} modified the original \citet{1972ApJ...176....1G} formulation to determine the stripping radius for spherically symmetric density distributions. Their stripping condition then reads analogously

\begin {equation}
\rho_{amb} \times v^2_{rel} > g_{max}(R)\times \Sigma_{gas}(R)
\end {equation}

where $\rho_{amb}$ here is the density of the surrounding halo material, $v_{rel}$ is the relative velocity between the cloud and the ambient medium, $g_{max}(R)$ is the maximum restoring acceleration in wind direction and $\Sigma_{gas}(R)$ is the surface density of the gas projected in wind direction. $R$ is the distance from the cloud center perpendicular to the direction of the ram pressure. For a schematic diagram see Fig. 3 of \citet{2008MNRAS.383..593M}. To calculate $R_{strip}$, where

\begin {equation}
\rho_{amb} \times v^2_{rel} = g_{max}(R_{strip}) \times \Sigma_{gas}(R_{strip})
\label{eq:r.strip}
\end {equation}

\noindent at any time, we use the 3D data cube and drill bars out of the cloud in the streaming direction of the halo gas with a cross section of one grid cell at the highest resolution and calculate the right-hand side directly from the gravitational potential and the gas density that is already stored in each grid cell. The ram pressure (left-hand side) would be constant for a standard windtunnel setup. In our case $\rho_{amb}$ stays constant over the simulation time but the relative velocity is increasing because of the gravitational attraction of the external mass. In our setups the velocity increases by about 6\% in 100 Myrs leading to an increase in the ram pressure by $\approx$ 12.4\%. Therefore the whole condition for RPS  shown in Eq.~\ref{eq:r.strip} is not constant in time because both the velocity $v_{rel}$, effecting the ram pressure term as well as the density distribution of the cloud changes, leading to a different column density distribution $\Sigma_{gas}(R)$ and subsequently to a different gravitational restoring force.

\subsection{\label{subsec:resolution}Numerical Resolution}

In numerical simulations reliable results should not depend on the spatial resolution but only on the included physics. To make sure that the results are valid for our setup we checked for convergence in performing the same simulation run for different levels of refinement. We expect differences as well in the apparent substructure of the cloud as in the overall dynamical properties for models with low  refinement levels, compared to runs with sufficiently small grid sizes. Especially, mass loss procedures as RPS and Kelvin-Helmholtz instabilities are highly sensitive to the physical size of a single cell. 
In addition the equipotential volume within the cloud, i.e. where the cloud gas is gravitationally bound, is getting smaller when the cloud is approaching the external center-of-mass and is therefore likely to be not properly resolved in low resolution runs.

\begin{figure}[htbp]
\begin{center}
	\includegraphics[width=\linewidth]{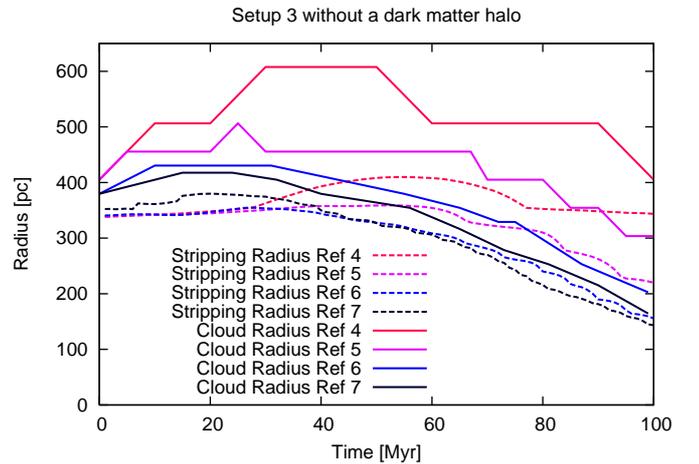}
\caption{ Comparison of the temporal evolution of cloud radius (full-drawn lines) vs. semi-analytically (Eq. \ref{eq:r.strip}) determined stripping radius (dashed lines) for Setup 3 without a DM halo and for different refinement levels 'Ref n'(for details see text). }
\label{fig:ramsetup3-wo}
\end{center}
\end{figure}

In Fig.~\ref{fig:ramsetup3-wo} the evolution of both the cloud radius and the stripping radius is shown for different levels of refinement. For the low resolution runs with $64^3$ cells (Ref 4) and $128^3$ cells (Ref 5) the calculated stripping radius suffers from the poorly resolved gravitational potential within the cloud. Therefore the maximum restoring acceleration $g_{max}$ is underestimated. For higher resolution runs with $256^3$ cells (Ref 6) and $512^3$ cells (Ref 7) results are converging. 
Comparing the analytically determined stripping radius to the cloud radius for the same resolution, it is evident that the evolution of the cloud radius follows the changes in the stripping radius. Therefore, dynamical stripping due to ram pressure is the dominant mass loss procedure in the considered scenario.

\subsection{\label{sec:other}Further mass-loss processes}

Even though we performed highly refined simulations with a resolution as low as 5 pc, it is not easily possible to take processes into account that are happening on sub-grid scale. In this section, we discuss the relative importance of other than the above described mass-loss processes, as namely Kelvin-Helmholtz (KH) instability (Sec.~\ref{sec:kh}), Rayleigh-Taylor (RT) instability (Sec.~\ref{sec:rt}), mass loss due to intrinsic velocity dispersion (Sec.~\ref{sec:veldisp}) and tidal disruption (Sec.~\ref{sec:tidaldisrup}).

\subsubsection{\label{sec:kh}Kelvin Helmholtz instability}

KH instability \citep[for detailed analysis, see][]{1961hhs..book.....C} can occur when a velocity shear is present within a continuous fluid or, when there is sufficient velocity difference across the interface between two fluids. It can be an important process that leads to a high mass loss due to so-called KH stripping of cloud material. Stability analysis of the KH instability shows a condition for the growth of perturbation in dependence on the wavenumber $k=2\pi/\lambda$. Perturbations with a wavenumber of $k \le k_{min}$ are damped while perturbations with $k > k_{min}$ are enhanced (see Eq.~\ref{eq:kmin}). 
The critical value $k_{min}$ can be derived from

\begin{equation}\label{eq:kmin}
k_{min} = \frac{g_z(\alpha_1-\alpha_2)}{\alpha_1 \alpha_2(v_1-v_2)^2}
\end{equation}

with $ \alpha_i = \rho_i/(\rho_1+\rho_2)$ (i=1,2), where $g_z$ is the gravitational acceleration perpendicular to the relative motion, $\rho_1$ and $\rho_2$ are the densities e.g. of cloud and hot gas at the edge of the cloud, and $(v_1-v_2)=v_{rel}$ is the relative velocity between the cloud and the ambient hot gas. For the initial values of setup 2-DD, $k_{min}=0.59\,kpc^{-1}$ which corresponds to a wavelength of $\lambda_{max}=10.8\,kpc$. Therefore all perturbations with a wavelength shorter than $\lambda_{max}$ are leading to KH instability. 

On the other hand, the KH timescale $\tau_{KH}$ can be estimated as the time after which the whole cloud mass $M_{gas}$ is removed from the cloud by the continuous mass loss due to KH instability. The mass-loss rate is given by \citet{1982MNRAS.198.1007N} with  $\dot{M}_{KH}=\pi r^2 \rho_{amb} \times v_{rel}$, where r is the cloud radius, $\rho_{amb}$ is the ambient density of the hot halo gas.  The KH timescale is given by:

\begin{equation}
\tau_{KH} = \frac{M_{gas}}{ \dot{M}_{KH}}
\end{equation}

Inserting typical values for our models, the KH timescale is $5.8\,\mbox{Gyrs}$, which is much longer than our simulation time of 100 Myrs. Starting at 50 kpc with a typical velocity of $200\kms$, the cloud reaches the galactic disk after 240 Myrs. Therefore, stripping by KH instability does not play a major role on the evolution of HVCs.

\subsubsection{\label{sec:rt}Rayleigh Taylor instability}

For an accelerated cloud with a higher density than the ambient medium also the conditions for a RT instability are fulfilled \citep{1961hhs..book.....C}. Stabilizing mechanisms which can suppress RT instabilities are magnetic fields \citep{1961hhs..book.....C}, heat conduction \citep{2007A&A...472..141V} and self-gravity. \citet{2008A&A...483..121R} investigated the conditions for RT instabilities in a self-gravitating cloud. They found that without magnetic fields, the expression for whether a RT instability is present at a radius R from the cloud center is equivalent to the expression for RPS. Subsequently, in regions where RT instabilities are suppressed the cloud is also stable against RPS, while in regions with a larger distance to the center, the gravitational restoring force is smaller which leads to both RPS and the occurrence of RT instabilities.

\subsubsection{\label{sec:veldisp}Velocity dispersion}

As shown in Fig.~\ref{fig:rochelobe2dd} a significant fraction of the cloud mass is located outside the Roche lobe at any time. Therefore, this gas is gravitationally unbound to the cloud and very likely to be lost within a few dynamical timescales $\tau_{dyn}$ due to the thermal pressure.
For a temperature of 1000 K (Setup 2, 2-D and 2-DD) the mean thermal velocity of the hydrogen atoms within the cloud is $\sigma_H=5\kms$ and for a temperature of 2000 K (Setup 3 and 3-D) $\sigma_H=7\kms$. The dynamical timescale, determined by $\tau_{dyn} = D_{gas}/\sigma_H$, where $D_{gas}=2R_{gas}$ is therefore between $\approx 50\,\mbox{Myrs}$ (Setup 2-D) and $\approx 105\,\mbox{Myrs}$ (Setup 3). This is of course only valid for regions outside the Roche lobe and for a constant temperature and cloud diameter over time, which is not realistic. Anyway it provides a first estimate that the simulation time of 100 Myrs is of the order of the dynamical timescale or twice of it and therefore this process is dominated by the ram pressure which is happening on a much shorter timescale.

Unless all other effects mentioned before, this mass-loss process is not only valid for the gaseous component but also for the DM subhalo that is surrounding the gas cloud. As a typical velocity for the DM minihalo \citet{1995ApJ...447L..25B} finds $v_0=17.7(r_0/\mbox{kpc})^{2/3}\kms$ while the virial radius as the extension of the halo is given by $R_{\rm vir}=3.4 \times r_0$ \citep{1995ApJ...447L..25B}. Subsequently the dynamical timescale for the DM component is $\tau_{dyn} \approx 210\,\mbox{Myrs}$ for the $r_0=150\,pc$ halo and $\tau_{dyn}\approx 250\, \mbox{Myrs}$ for the $r_0=250\,pc$ halo. This is in both cases more than twice the simulation time, which allows us to use a DM halo that is static with respect to the cloud center and is therefore accelerated together with the gaseous component in the direction of the external mass.

\subsubsection{\label{sec:tidaldisrup}Tidal disruption}

On their journey to the galactic disk of their host galaxy, high velocity clouds would not only be ram pressure stripped but also tidally stretched in the gravitational field of the galaxy. According to \citet{1999ApJ...514..818B} the already disrupted clouds would correspond to the large HVC complexes with large angular extents and distances in the order of a few kpc. A useful estimate for the cloud's tidal radius is provided by the Jacobian limit \citep{1988galdyn} $r_J$:

\begin{equation}
   r_J = \pm D \left [ \frac{m}{M(3+m/M)}\right ]^{1/3}
\end{equation}

\noindent where $D$ is the distance between the two mass centers, $m$ is the HVC mass and $M$ the galaxy mass. Initially at a distance of 50 kpc all cloud radii lie well within the Jacobian limit. For Setup 2-DD for example the Jacobian limit is 4.5 times larger than the cloud radius without a DM subhalo, 10.3 times with a $r=150\,pc$ halo, or 15 times with a $r=250\,pc$ halo. After the simulation time of 100 $Myrs$ and even with an assumed extreme mass loss of 50\% the Jacobian limit is still between 2 and 7 times larger than the initial cloud radius. In addition, also the virial radius of the DM minihalo remains always within the Jacobian limit. Therefore we conclude that tidal disruption does not play a role on radial orbits and at these distances from the galactic center.

\section{\label{sec:results}Results}

\subsection{\label{sec:HT}Head-tail structure}

For all simulation runs without DM it is possible to reproduce the observed and expected head-tail structure in the density distribution of the cloud. For a better comparison with observations, the 3D density is integrated in z-direction in order to get a 2D column density distribution. 
Iso-density contours are shown in Fig.~\ref{fig:rochelobe2dd} as well as in Fig.~\ref{fig:nature2dd} and Fig.~\ref{fig:nature3}. While in runs without an additional DM potential the iso-density contours are distorted even close to the cloud's mass center, they stay very circular in models in which the cloud is embedded in a DM subhalo. 
A remarkable difference not only in the head-tail structure between simulations with and without DM can be found in the results of Setup 2-DD (Fig.~\ref{fig:nature2dd}) where the cloud without DM is totally disrupted while there is hardly any difference in the column density distribution between the runs with DM halo. We conclude that while the existence of a DM halo has a large impact on the head-tail structure of the HVCs, it is not very sensitive to the parameters of the DM halo as $M_{DM}$, $r_0$ and whether the DM density profile has a core or a cusp in the center (see Sec.~\ref{sec:dmshape}).

\subsection{\label{sec:lineofsight}Line-of-Sight Effects}

Because the HVC motion is by far not radially towards the observer, but might happen ballistically, we have chosen different line-of-sight angles and integrated the radial velocity components from a virtual observer. To create a plot in the velocity-position space not only the velocity is affected by the line-of-sight effect but also the spatial coordinate (see. Fig. \ref{fig:setup3_contourvel} for further explanations).

We find that different velocity gradients along the head-tail structure of the gas cloud can be easily created only by changing the inclination of the line-of-sight aspect angle. For inclinations of 15$\deg$ and 30$\deg$ also changes in the velocity gradient appear, as observed in different HVCs. These changes appear in Setup 3 at distances from the cloud center of about 300 pc which is far outside the Roche lobe. Therefore the ``knees" in phase space according to the different velocity gradients are not directly connected to the Roche lobe, which is at a distance of less than 100 pc (Setup 3) after 100 Myrs of simulation time.

For comparison between our models and HVCs observations \citep[e.g. by][] {2000A&A...357..120B, 2001A&A...370L..26B, 2006A&A...457..917B} the head-tail structure as a simple horizontal cut through the HVCs is not an appropriate measure to decide about the existence of a DM subhalo. In the present paper, we limit ourselves to only radially infalling HVCs, while we suggest that inclined HVC motions -- and most probably with a significant perpendicular velocity component -- could lead to the observed knee in the velocity-position distribution.

\begin{figure}[htbp]
\begin{center}
	\includegraphics[width=\linewidth]{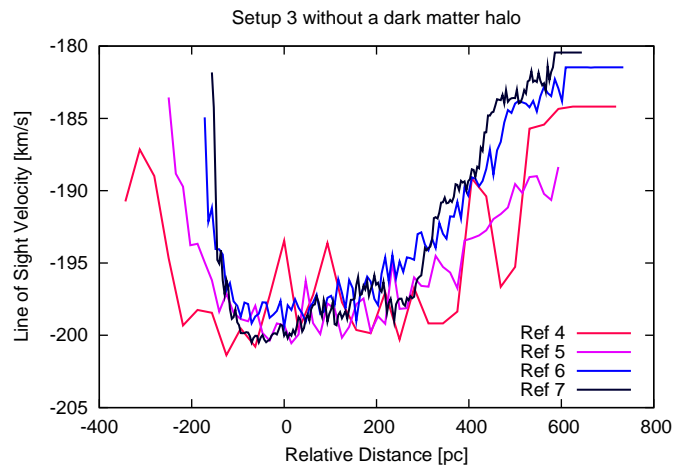}
\caption{The average line of sight velocity under an inclination angle of $\alpha=30\deg$ in each position bin after 100 Myrs of simulation time for four different levels of refinement.}
\label{fig:lossetup3-wo}
\end{center}
\end{figure}

Since we investigate the location of the HVC in velocity-position phase space, we check also the final model for convergence with runs on higher resolution. The line-of-sight dependent contour plots shown in Fig.~\ref{fig:setup3_contourvel} are analyzed in more detail for runs with different refinement levels. The mean velocity is calculated in bins of the initial cell size for each refinement and shown for an angle of $\alpha=30\deg$ in Fig.~\ref{fig:lossetup3-wo} for refinement levels 4, 5, 6 and 7. While the gradient of the velocity in the tail part is very different between refinement runs 4, 5 and 6 the result is nicely converging between refinement 6 and 7.  
 
 To compare all simulation runs for their behavior in phase space for different line-of-sight directions and independency on their DM content the plots for Setup 2, Setup 2-D, Setup 3 and Setup 3-D are shown in Fig.~\ref{fig:set2_red}, Fig.~\ref{fig:set2D_red}, Fig.~\ref{fig:set3_red} and Fig.~\ref{fig:set3D_red}.

\subsection{\label{sec:parameter}Parameter Study}

\subsubsection{\label{sec:windvel}Effect of the initial wind velocity}

While all models mentioned before start with an initial wind velocity of $200\kms$ i.e. supersonic motion with respect to the ambient medium of Mach 1.7, in this section, we present models of Setup 2-DD without DM halo and with a DM halo of $r_0=250\,pc$ at an initial velocity of $250\kms$ (Mach 2.1) in order to investigate the effects caused by different relative velocities $v_{rel}$. 

\begin{figure}[htbp]
\begin{center}
	\includegraphics[width=\linewidth]{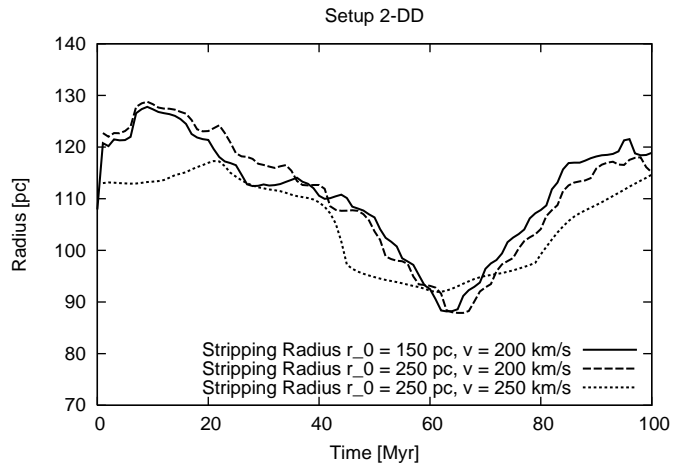}
\caption{Semi-analytically derived stripping radius for Setup 2-DD with DM minihalos of $r_0 = 150\,pc$ (full-drawn line, Run VI) and $r_0 = 250\,pc$ for an initial relative velocity of $200\,km\,s^{-1}$ (dashed line, Run IX) and  $250\,km\,s^{-1}$ (dotted line, Run XI).}
\label{fig:ramsetup2-DD}
\end{center}
\end{figure}

The gas clouds, however, are not directly experiencing the full wind speed due to a quickly forming bow shock which leads to a post-shock region with subsonic velocity. Fig.~\ref{fig:ramsetup2-DD} demonstrates that only a small difference in the evolution of the stripping radii exists between the runs with the different velocities.

\subsubsection{\label{sec:dmmass}The dark matter mass}

Setup 2-DD represents a cloud that is in hydrostatic equilibrium under its total mass, consisting of the gas ($M_{gas} = 4.4\times 10^5\,M_{\odot}$) and a DM minihalo with a scale radius of $r_0=250\,pc$ and a mass of $M_{DM}=1.63\times10^7 \Msun$.  Simulations with DM minihalos of $r_0 = 150\,pc$ and $r_0 = 250\,pc$ are performed as well as one model simulation with a pure gas cloud only (see Fig. \ref{fig:nature2dd}). The maximum refinement level for all three runs is set to 7, leading to an effective resolution of 5 pc. (For further informations about Setup 2-DD see Tab.~\ref{tab:overview}.)
While in the run without DM the cloud is disrupted and significantly decelerated after not more than 70 Myrs, both clouds with the Burkert DM halo remain almost unaffected by ram pressure with only small mass loss. They do not show significant differences, even though the total DM mass with $r_0 = 250\, pc$ is more than three times higher than that of the halo with $r_0 = 150\, pc$.

The time evolution of the model without a DM subhalo is shown by its column density distribution in Fig.~\ref{fig:time_setup2-DD} (Appendix). Until 50 Myrs the ram pressure affects only the outer HVC shells while the main HVC body remains untouched. 
In this run the central part of the gas cloud is highly compressed by means of the strong drag force and (starting after 60 Myrs) the peak density is shifted away from the simulation rest frame, i.e. is decelerated. 

Due to the very high pressure in the head part of the cloud the gas starts to re-expand (perpendicularly to the cloud's motion according to the Bernoulli effect) and to form a bow-like structure in front of its mass center (between 70 and 100 Myrs). 
This feature shows a compressed leading edge but low density in the  region behind. Its relatively large extent makes it very likely to be disrupted by both Rayleigh-Taylor instability and RPS in a short time. 
 
The highest gas mass fraction in our models is applied to Setup 3, where the DM mass is only 1.6 times larger than the gas mass. Even this light DM halo has a large impact on the stability of the cloud. While in the simulation run without DM the extent of the HVC perpendicular to the wind direction has decreased remarkably within 100 Myrs, the cloud with the surrounding DM halo conserves the initially spherical-symmetric shape much better (see Fig.~\ref{fig:nature3}). The mass loss after 100 Myrs is for both runs, with and without DM, of the order of a few percent. In Setup 3 without a DM halo, the ratio between bound mass and total mass equals 0.966 which is means that the mass loss is twice as large as in the simulation run with a DM halo, where the fraction of bound mass is 98.31 \%. This means that the self-gravity of a gas cloud of $3.1\times10^6 \Msun$ alone is strong enough to keep most of the material bound in the first 100 Myrs and cannot be neglected in simulations.

We determined the bound mass by comparing the sum of kinetic and internal energy to the corrected potential energy for every cell within the computational domain. Because we are interested in the binding energy to the cloud or cloud-halo system rather than to the external point mass, the total gravitational potential is reduced by the potential which is produced by the host galaxy. Every cell in which the binding energy dominates contributes to the bound mass. 

For Setup 2-DD without a DM halo (Run VI), 95.72\% of the initial cloud mass is still bound after 100 Myrs. The same setup but embedded in an $r_0 = 150\,pc$ (Run VII) or $r_0 = 250\, pc$ (Run IX) halo reduces the amount of mass loss by a factor of 7 - 8 and leads to a bound mass of 99.39\% ($r_0 = 150\,pc$) and 99.49\% ($r_0 = 250\, pc$) of the initial gas mass after 100 Myrs.

\subsubsection{\label{sec:dmshape}The dark matter density profile}

The DM density distribution especially in low mass system is still under discussion \citep[e.g.][]{2007MNRAS.381.1450N, 2012A&amp;A...540A..70R, 2010MNRAS.402...21N, 2012MNRAS.419..184A, 2011ApJ...742...20W, 2012ApJ...751L...5T, 2012MNRAS.423.2190S}. 

To provide a full sample, we started an additional run with an NFW \citep{1997ApJ...490..493N} DM density profile to numerically investigate possible differences in the ram pressure stripping.

We re-simulated Run IX, with an NFW halo which shows a cuspy density profile unlike the centrally flat Burkert halo. We decided to set up an NFW halo with the same 
total DM mass as in Run IX ($M_{200} = 1.63 \times 10^7\, M_{\odot} $) so that we can compare the differences that appear due to the shape of the DM halo.
The virial radius of this NFW halo $r_{200} = 5.5\, kpc$ and the circular velocity at $r_{200}$ is: $v_{200} = 3.57\, km\, s^{-1}$. 
For a full description of the DM density profile:

\begin{equation}
\rho_{NFW}(r) = \frac {\delta_c \rho_c}{\frac{r}{R_s}\left ( 1 + \frac{r}{R_s} \right)^2 }
\end{equation}

\noindent with $\delta_c$ the overdensity relative to the critical density of the universe $\rho_c = 3H^2 / (8\pi G)$ (with $H=65\,km\,s^{-1}\,Mpc^{-1}$), the concentration parameter c in addition to the virial mass $M_{200}$ is needed. This parameter differs significantly from object to object and doesn't show a clear correlation with the mass of the object. Fitting rotation curves from galaxies from the THINGS survey shows concentration parameter in large varieties  \citep{2011AJ....142..109C}
and vary between $c = [1 \cdots 30]$ for dwarf galaxies \citep{2001MNRAS.325.1017V}. 

Using a concentration parameter of c = 22, which lies well within the measured range, leads to a scale radius of $R_s = r_{200} / c = 0.25 \, kpc$, which is the same as in the comparison halo with a Burkert profile. 
With the concentration parameter c, the overdensity $\delta_c$ is determined by:

\begin{equation}
\delta_c = \frac{200}{3} \frac{c^3}{[ln(1+c)- c/(1+c)]}
\end{equation}

\begin{figure}[htbp]
\begin{center}
	\includegraphics*[width = \linewidth] {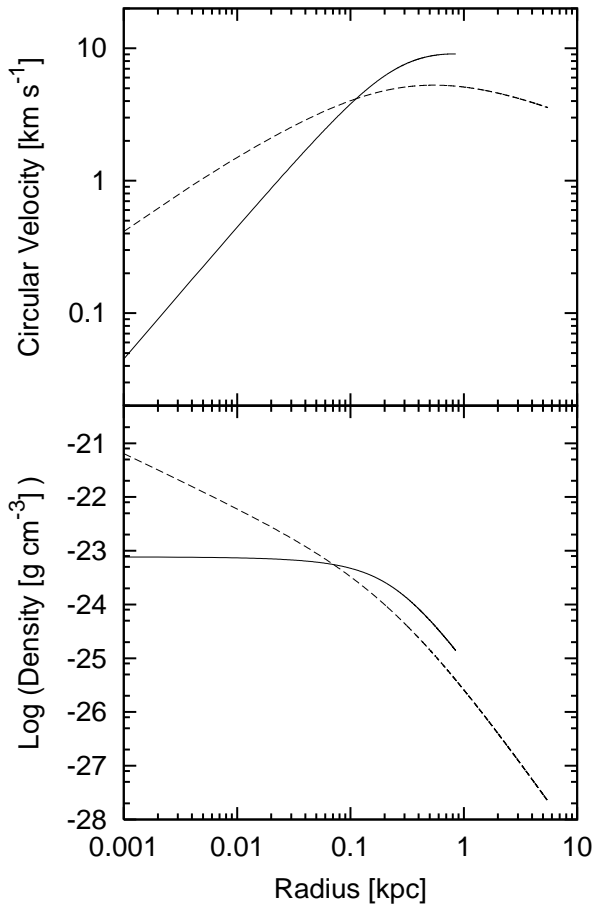}
\caption{Comparison of the DM density profile (bottom) and the circular velocity profile (top) for DM halos with a scale radius of 250 pc and a mass of $1.63 \times 10^7\,M_{\odot}$. The solid line shows the Burkert profile (central core profile), while the dashed line represents an NFW profile with a cusp in the center.}
\label{burk_nfw}
\end{center}
\end{figure}

\noindent The radial distribution of the DM density as well as the circular velocity of both the Burkert and the NFW halo are shown in Fig.~\ref{burk_nfw}.

The results (see Fig.~\ref{fig:nature2dd}) demonstrate that the final column density distribution and therefore the existence of a head-tail structure is neither highly dependent on the DM mass (see also Sec.~\ref{sec:dmmass}) nor on the shape of the DM potential.

\section{Conclusions}

Summarizing, the following major issues from the presented series of model runs can be derived. While it is possible to reproduce the observed change in the velocity gradient, this feature is not directly connected to the Roche lobe and therefore the position of the observed ``knee" in velocity-position phase space cannot help to distinguish between DM dominated and pure gas clouds. For a variety of initial distributions we can create these phase space distributions by changing the inclination of the line-of-sight. In a further study, clouds with a significant azimuthal motion must be tested with respect to their velocity gradient of the stripped gas for which we expect a more significantly observable velocity ``knee".

We simulate DM-embedded clouds with baryonic to total mass ratios between 0.03 and 0.39 and found that the results depend only marginally on the DM mass and profile and that already a DM halo with a mass of 1.6 times the baryonic mass stabilizes the gas cloud against ram-pressure disruption.

After 100 Myrs the column-density distributions of the clouds with DM halo can be well distinguished from pure gas clouds. While the isodensity lines of the core of gas clouds without an additional DM gravitational potential are losing their initial shape after half the simulation time and show a lot of substructures, for all DM-dominated models HVCs remain spherical and centrally concentrated, even if the DM content is low as in Setup 3.  Furthermore an extended head-tail structure at galactic heights between 30 and 50 kpc can only be reproduced for DM-free clouds.

In our simulations, clouds with DM - with baryonic masses in the range from a few times 10$^5$ to a few times $10^6 \Msun$ - approach the galactic disk like bullets with high velocities, while DM-free clouds are already decelerated and distorted at higher galactic heights. Observed HVCs with distances up to 30 kpc, like e.g. Complexes C and H, reveal a very heterogeneous multi-phase internal gas structure without any central concentration, which we only obtain in simulations without DM.

Furthermore, the observational lack of HVCs close to the Milky Way disk (only LVCs exist) and also that high-speed infalling \HI gas is not observed in galaxies with extended \HI envelopes, as e.g. NGC~2403 \citep{2002AJ....123.3124F} let us conclude that HVCs are decelerated to LVCs or disrupted on their infall into galactic halos, contrary to the simulated ``bullets" of clouds which have an accompanying DM minihalo.

Our DM-free models can both explain the proximity of the large HVC complexes to the galactic disk even though they start their journey from a height of 50 kpc above the plane, as well as their morphologies, like e.g. from elongated shapes for more massive clouds to head-tail structures and the inherent amount of substructures.
	
Our conclusions contain the following further serious consequences for the cosmological model and for our understanding of cosmological structure formation. 

\begin{enumerate}
	\item	 Since HVCs are DM-free, they cannot be the cosmological small-scale relics expected from $\Lambda$CDM structure.
	\item Since HVCs have typical distances to the disk of their host galaxy of not more than 50 - 60 kpc and do not populate the intergalactic space frequently enough, it is obtruding that, if they are DM-free, they originate from stripped-off gas of satellite galaxies or from condensations of diffuse intergalactic medium due to thermal instability rather than from an cosmological background.

\end{enumerate}


\begin{acknowledgements}
The authors acknowledge support with FLASH by Elke R\"odiger and Nigel Mitchell and helpful discussions with Bastian Arnold and Simone Recchi. The software used in this work was in part developed by the DOE-supported ASC / Alliance Center for Astrophysical Thermonuclear Flashes at the University of Chicago. 
Part of this study was funded by the project no. HE1487/36 (S.P.) within the DFG Priority Program ``Witnesses of Cosmic History: Formation and evolution of galaxies, their central black holes, and their environment". The numerical simulations are performed at the HPC astro-cluster of the Institute of Astronomy and at the Vienna Scientific Cluster (VSC1) under project no. 70128. The publication is supported by the Austrian Science Fund (FWF). The authors gratefully acknowledge helpful comments by the referee that lead to further clarifications.
\end{acknowledgements}

\bibliographystyle{aa}
\bibliography{Ploeckinger_Hensler_2012}

\newpage
\begin{appendix}
\section{}

\begin{figure*}[h]
\begin{center}
 	\begin {minipage}[b]{0.45\linewidth}
		\includegraphics[width=\linewidth, bb = 0 0 566 255,clip]{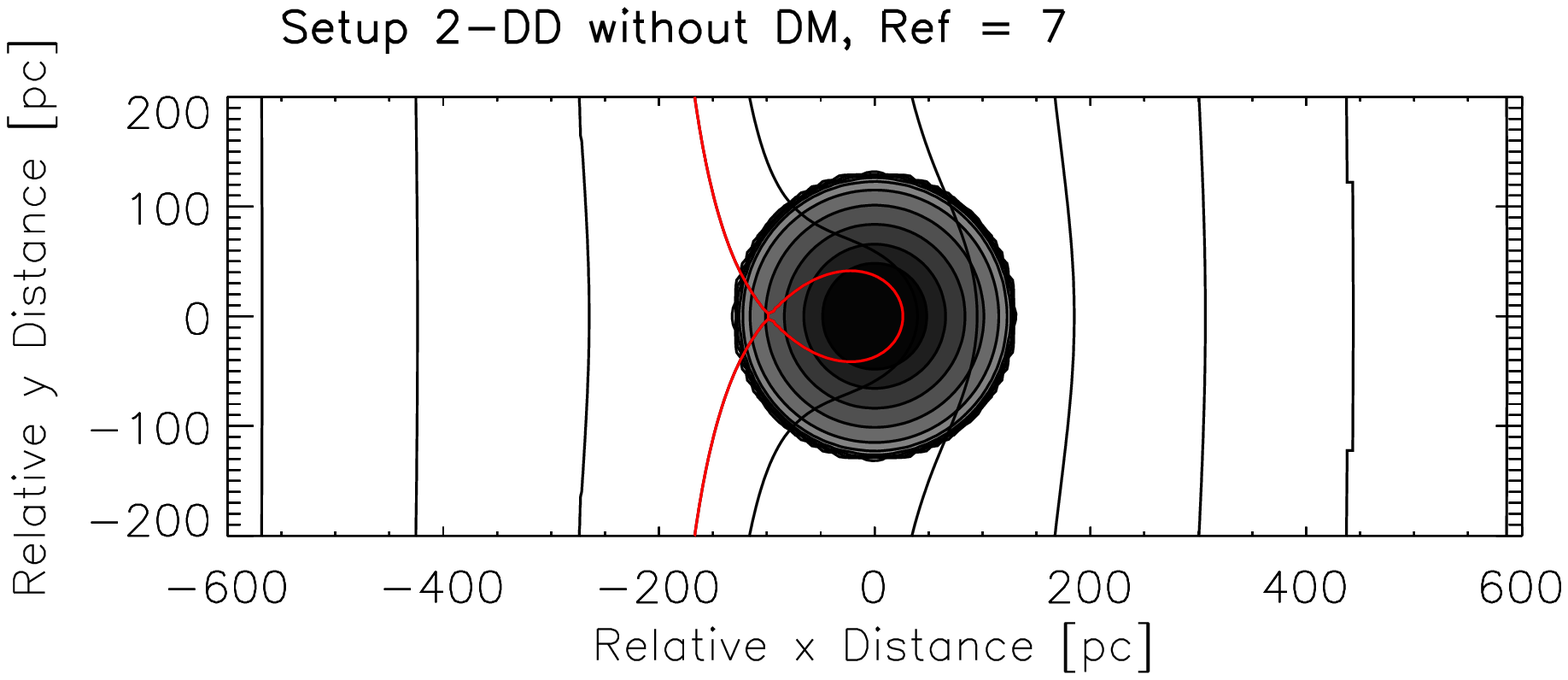}
	\end {minipage}
	 \begin {minipage}[b]{0.45\linewidth}
	 	\includegraphics[width=\linewidth, bb = 0 0 566 255,clip]{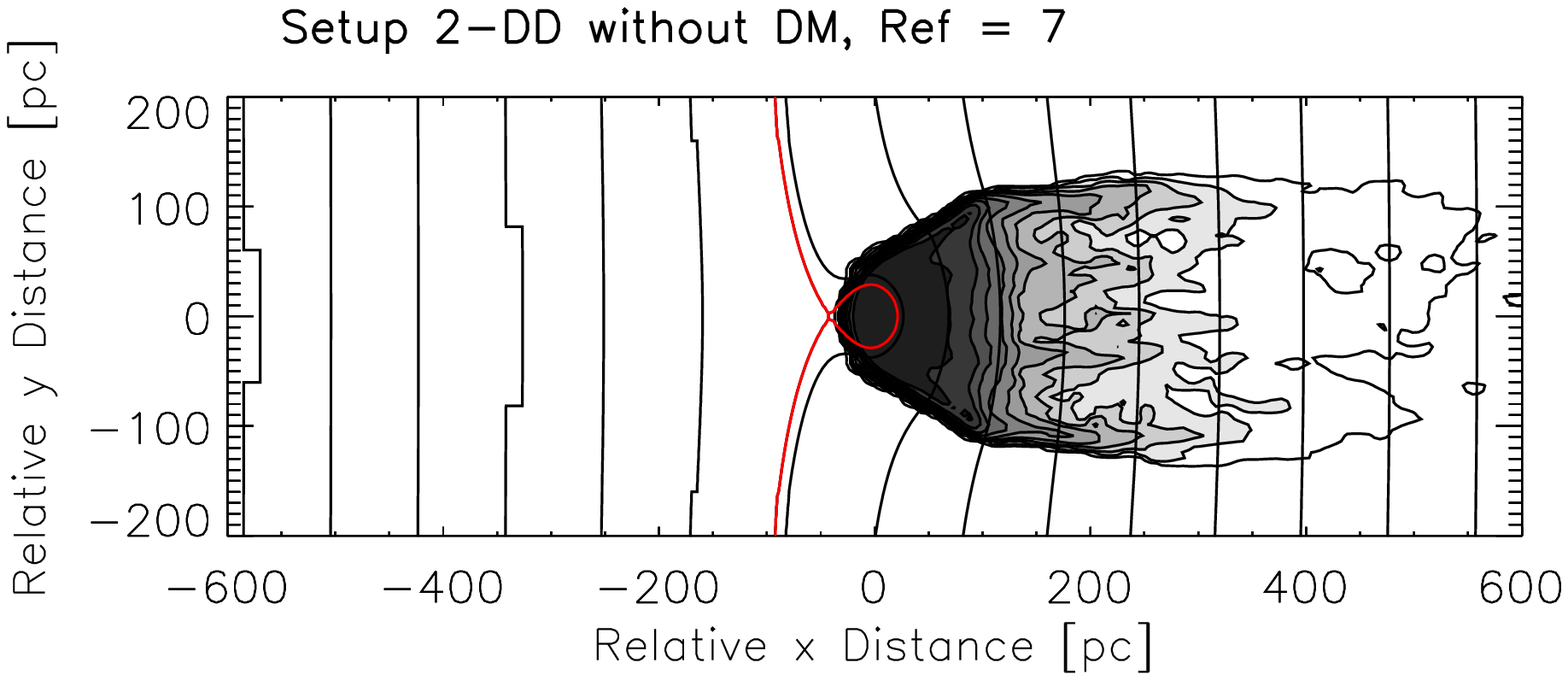}
	\end {minipage}
	
	 \begin {minipage}[b]{0.45\linewidth}
	 	\includegraphics[width=\linewidth, bb = 0 0 566 255,clip]{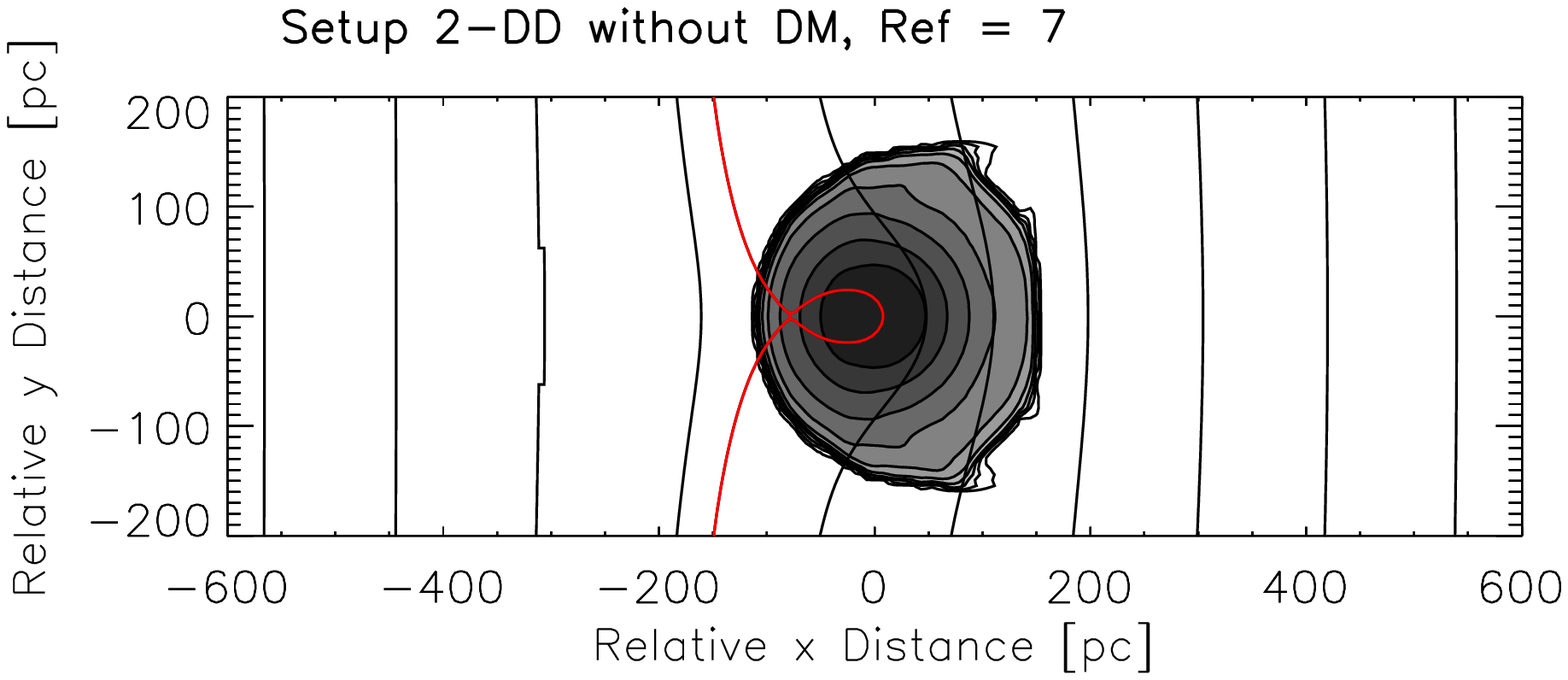}
	\end {minipage}
	 \begin {minipage}[b]{0.45\linewidth}
	 	\includegraphics[width=\linewidth, bb = 0 0 566 255,clip]{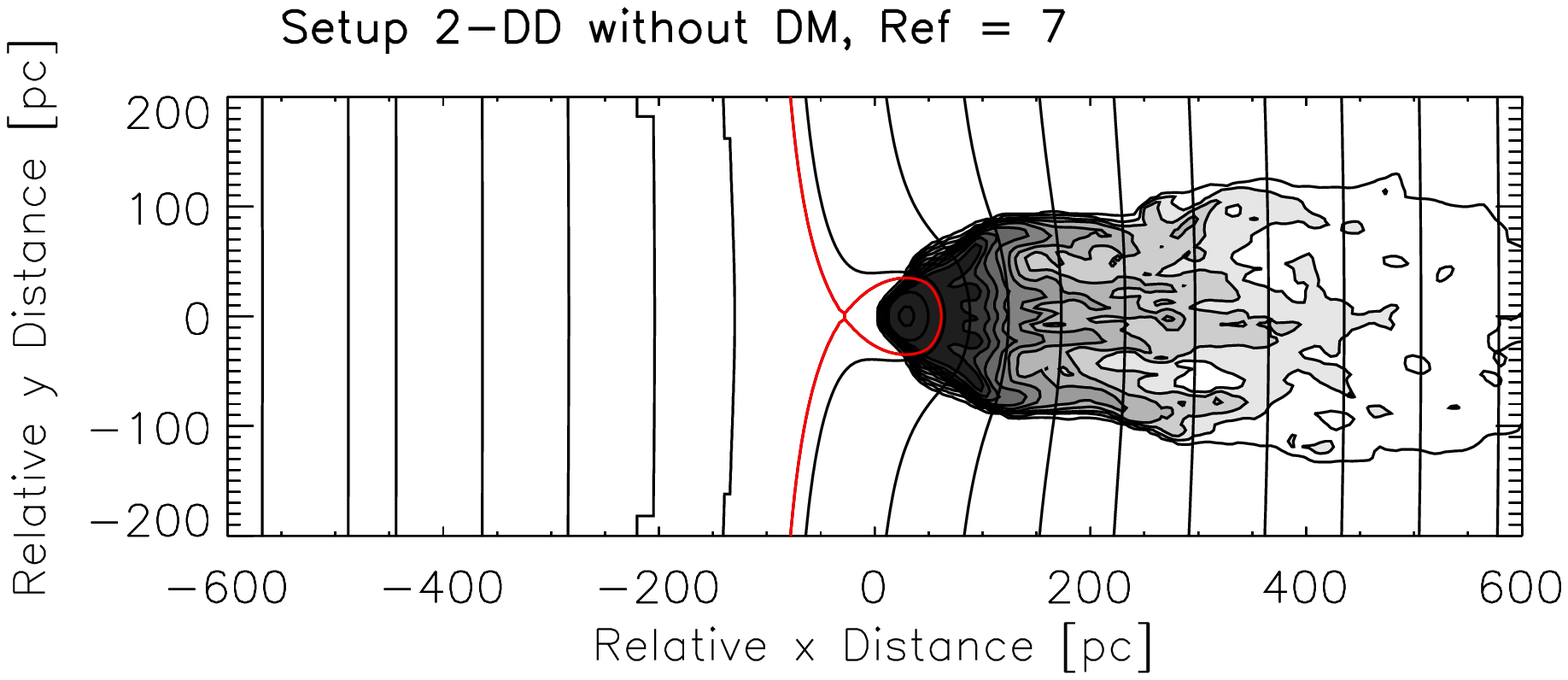}
	\end {minipage} 	
	
	\begin {minipage}[b]{0.45\linewidth}
		\includegraphics[width=\linewidth, bb = 0 0 566 255,clip]{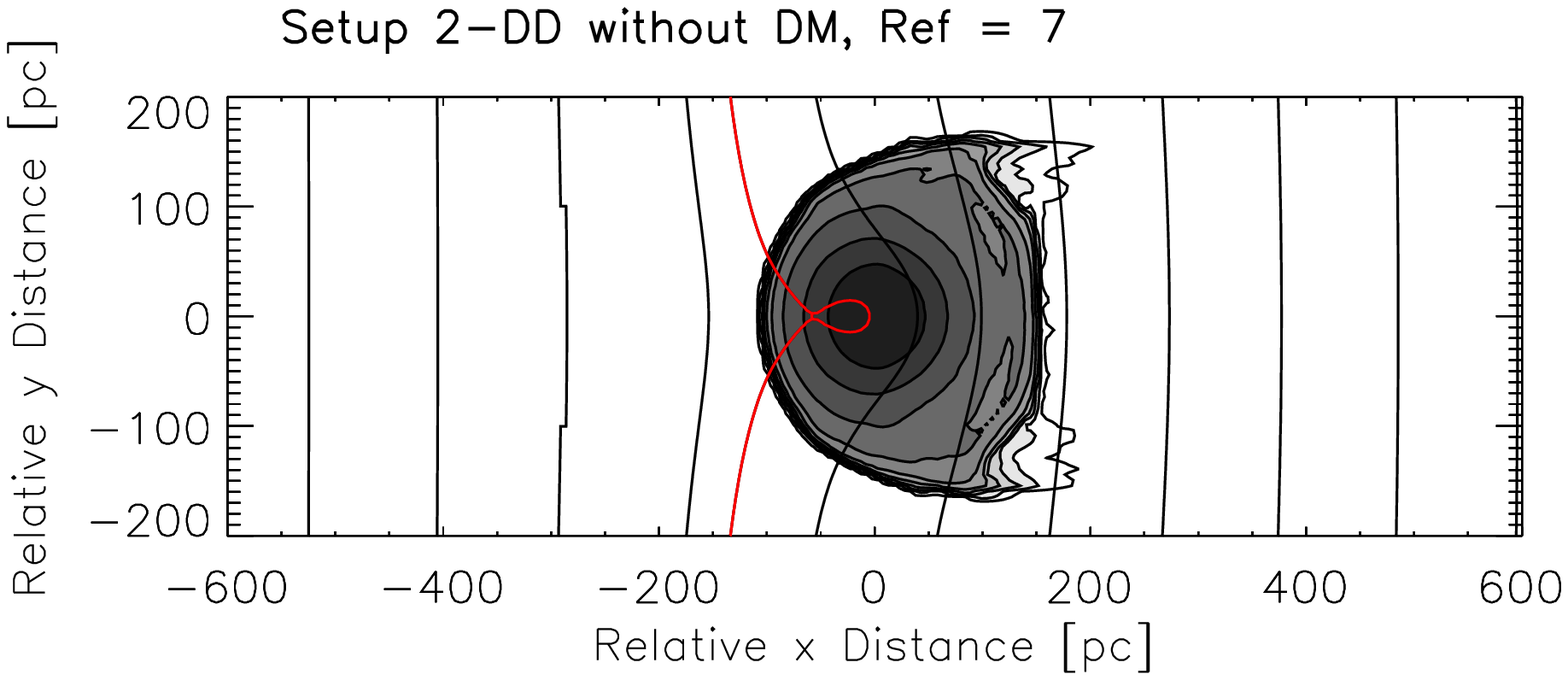}
	\end {minipage}
	 \begin {minipage}[b]{0.45\linewidth}
	 	\includegraphics[width=\linewidth, bb = 0 0 566 255,clip]{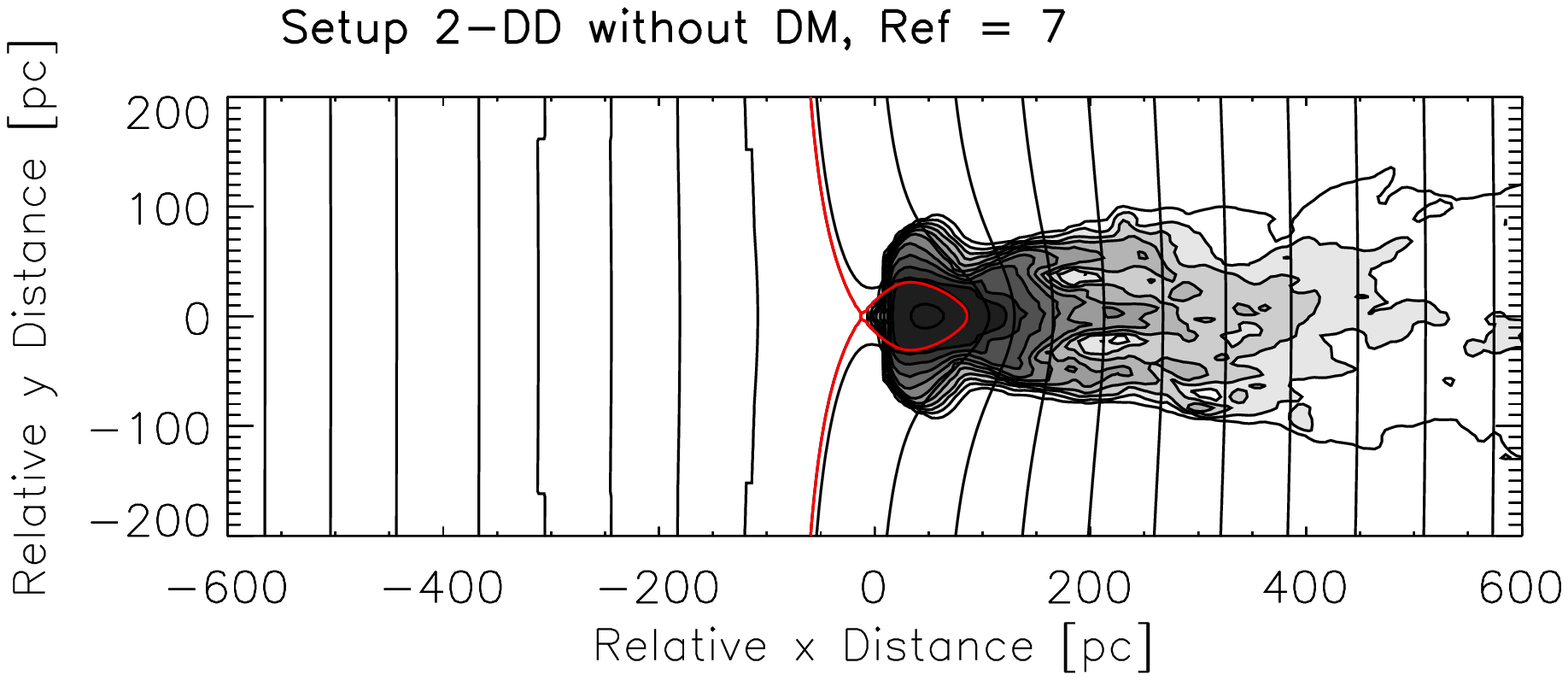}
	\end {minipage} 	
	
	\begin {minipage}[b]{0.45\linewidth}
		\includegraphics[width=\linewidth, bb = 0 0 566 255,clip]{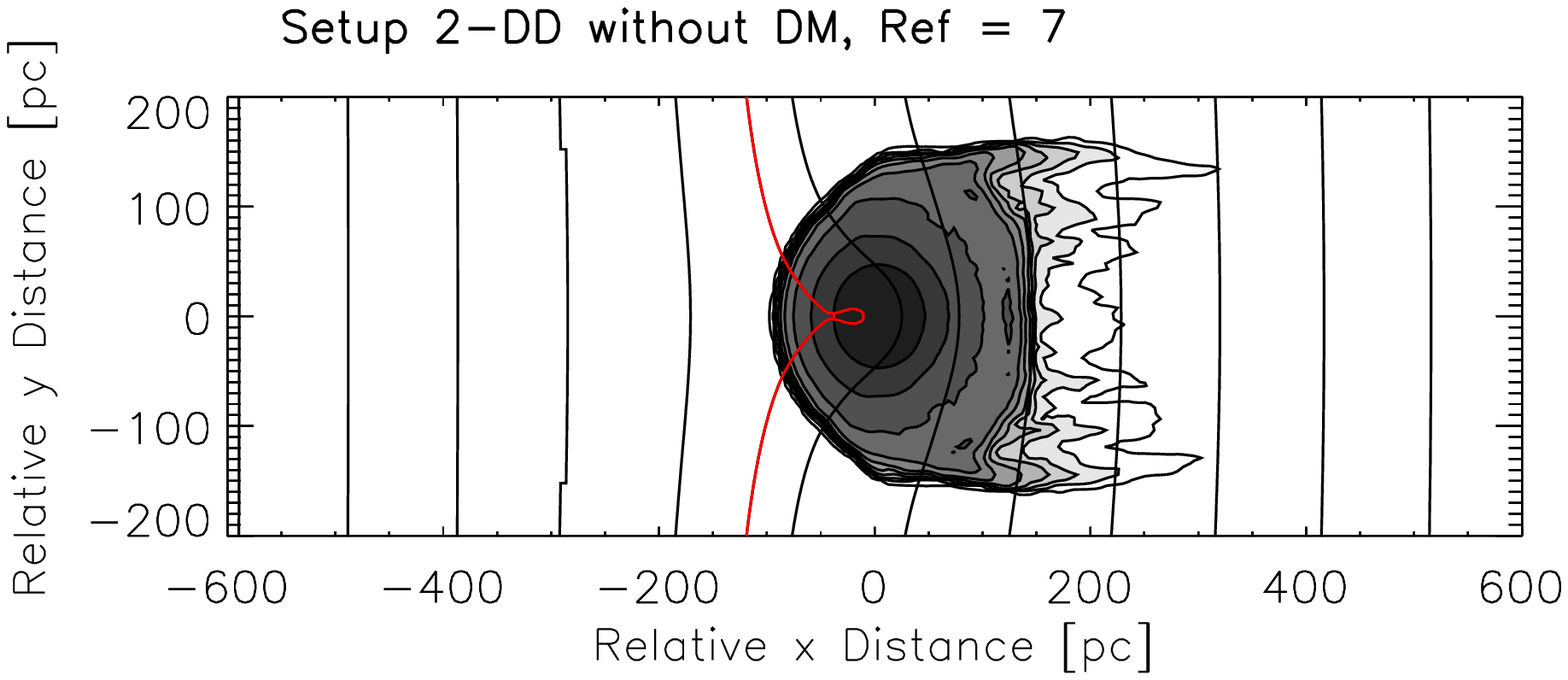}
	\end {minipage}
	 \begin {minipage}[b]{0.45\linewidth}
	 	\includegraphics[width=\linewidth, bb = 0 0 566 255,clip]{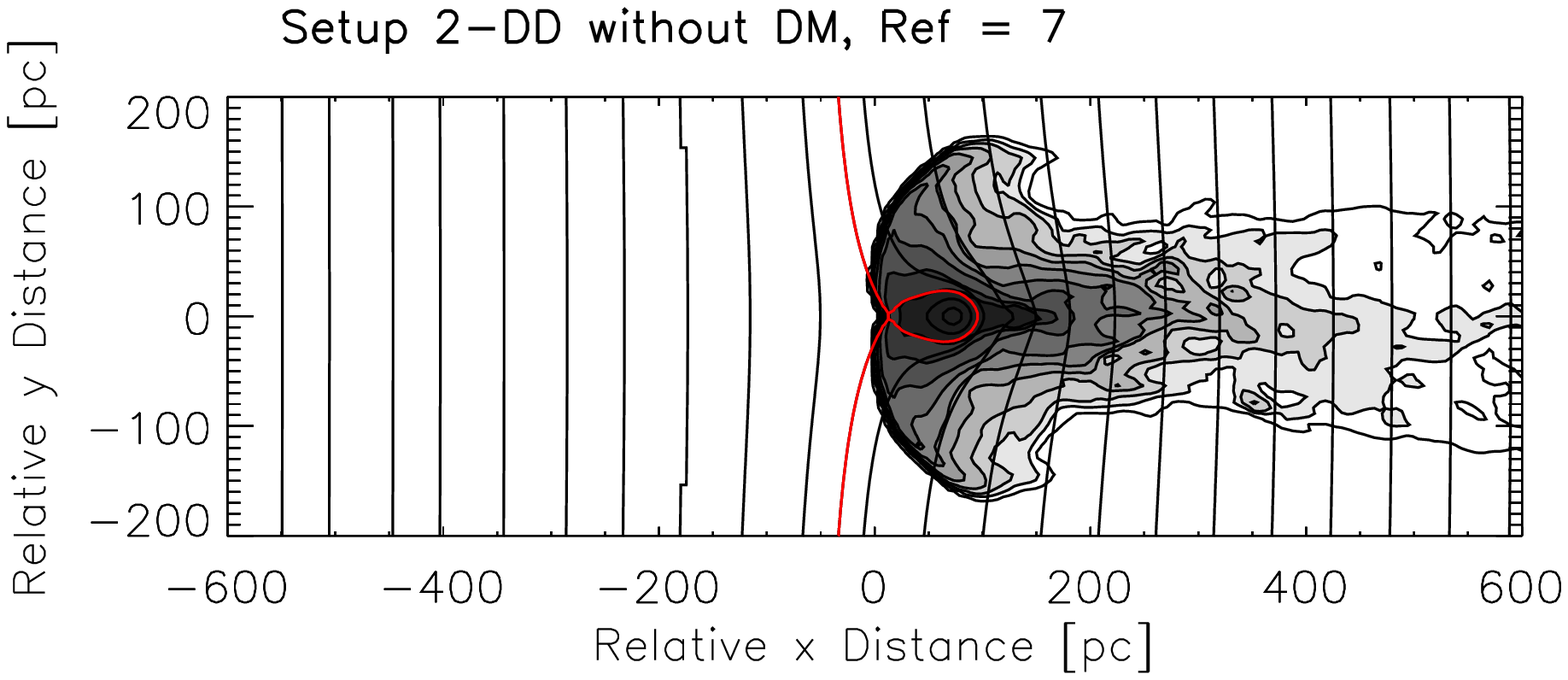}
	\end {minipage} 	

	\begin {minipage}[b]{0.45\linewidth}
		\includegraphics[width=\linewidth, bb = 0 0 566 255,clip]{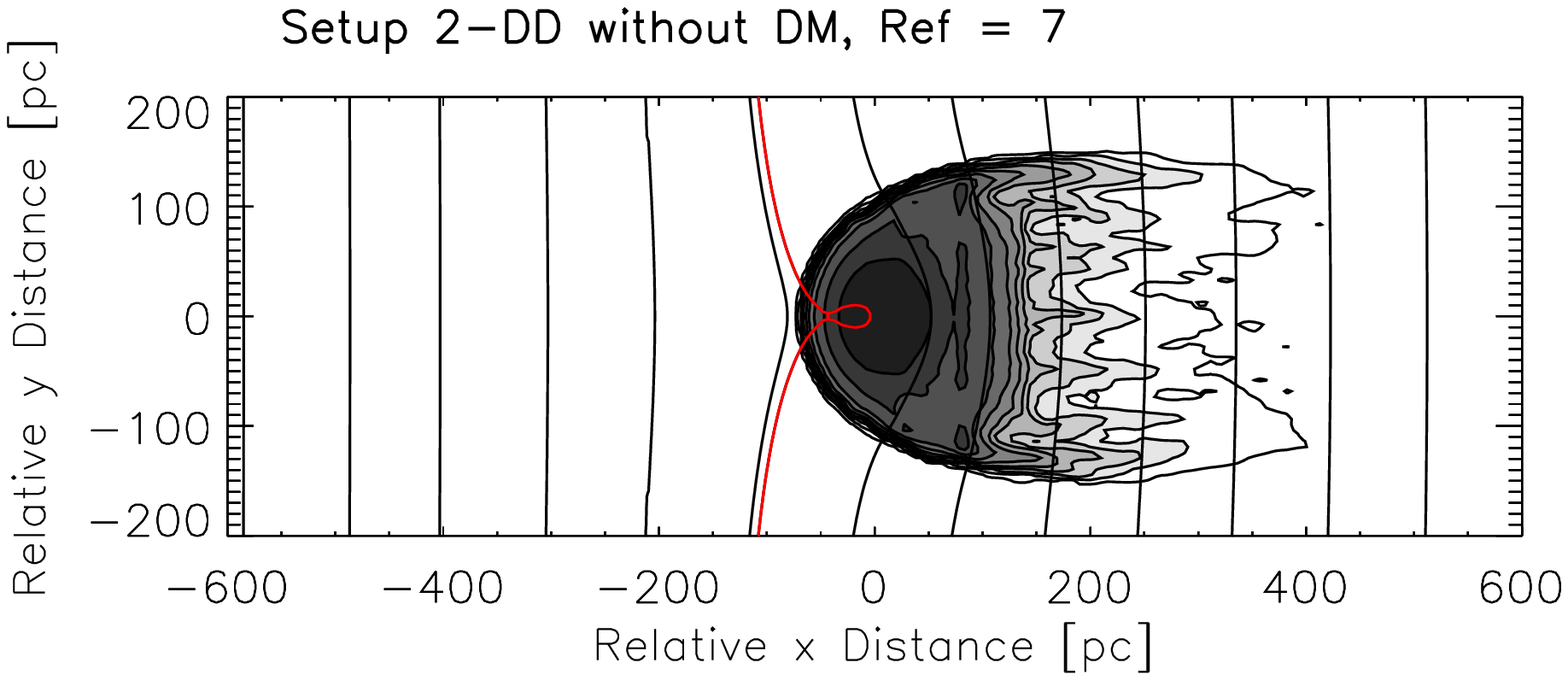}
	\end {minipage}
	 \begin {minipage}[b]{0.45\linewidth}
		\includegraphics[width=\linewidth, bb = 0 0 566 255,clip]{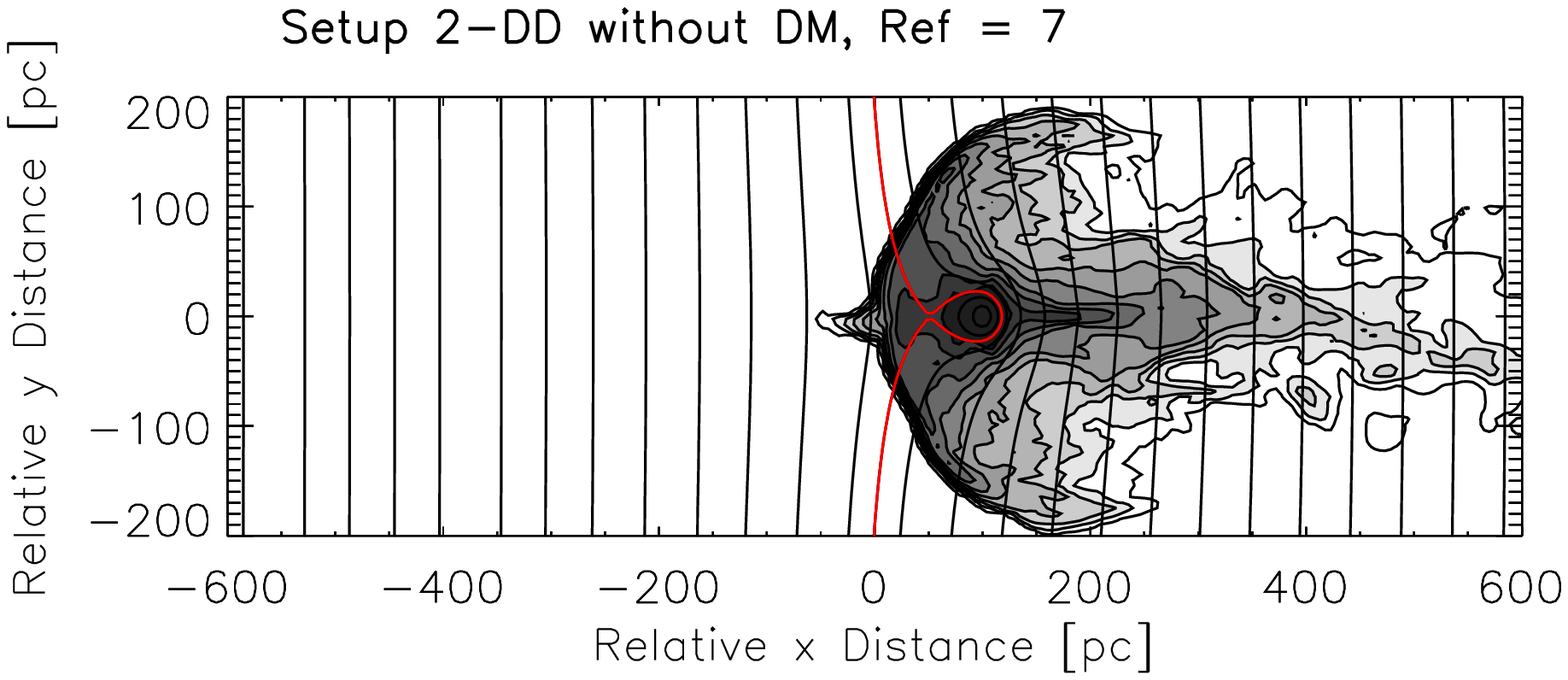}
	\end {minipage} 	

\caption{{\bf Setup 2-DD.} Time evolution of the column density and the equipotential distribution. Contours as in Fig~\ref{fig:rochelobe2dd}. From top to bottom, on the left-hand side the panels show snapshots at 0, 20, 30, 40 and 50 Myrs and on the right-hand at 60, 70, 80, 90 and 100 Myrs after the simulation onset.}
\label{fig:time_setup2-DD}
\end{center}
\end{figure*}

\begin{figure*}[htbp]
\begin{center}
	\includegraphics[width=\linewidth]{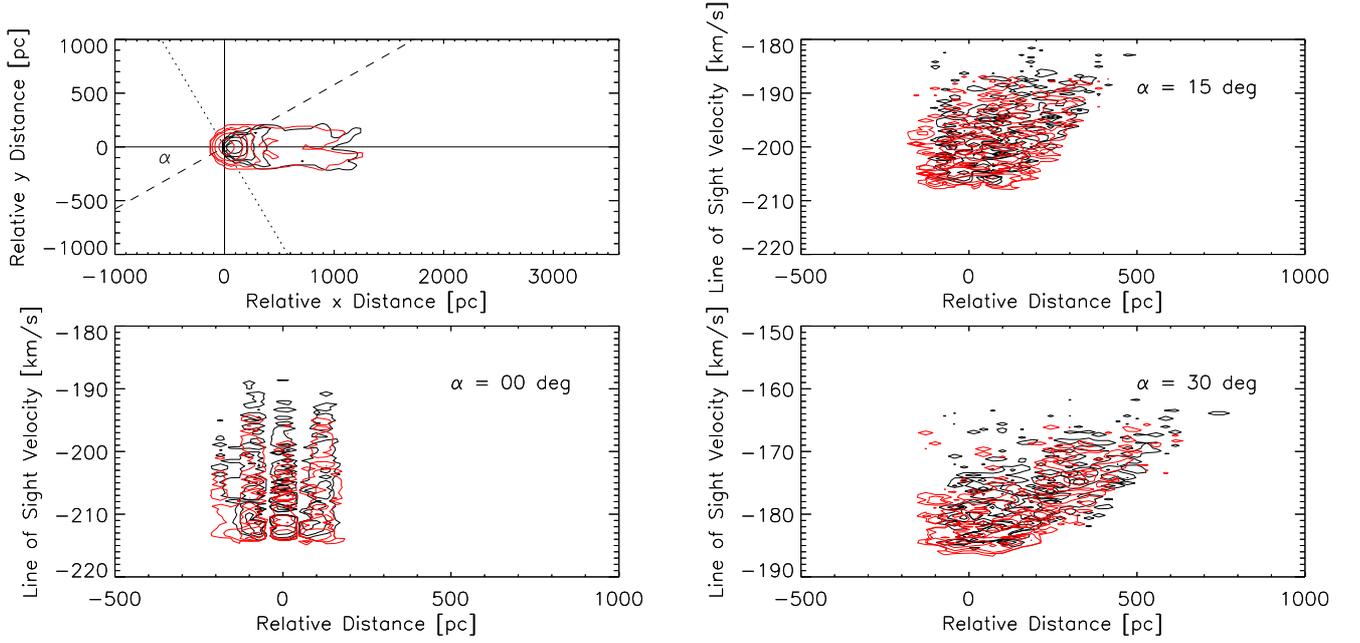}
\caption{{\bf Setup 2} with (red contour lines) and without a DM halo (black contour lines) at a maximum refinement level of 6. In the top left panel the contour lines represent the column density distribution $\log(N_{HI}) = 18.8\,...\,21.2\, H cm^{-2}$ in 0.6 dex after 100 Myrs of simulation time. For the line of sight plots in the panels on the right side and the panel at the bottom left, contour lines are at $\log(N_{HI}) = 18,\,19,\,20 \mbox{ and }21\, H\,cm^{-2}$. The angle $\alpha$ determines the inclination of the line of sight, for details see caption of Fig.~\ref{fig:setup3_contourvel}. }
\label{fig:set2_red}
\end{center}
\end{figure*}

\begin{figure*}[htbp]
\begin{center}
	\includegraphics[width=\linewidth]{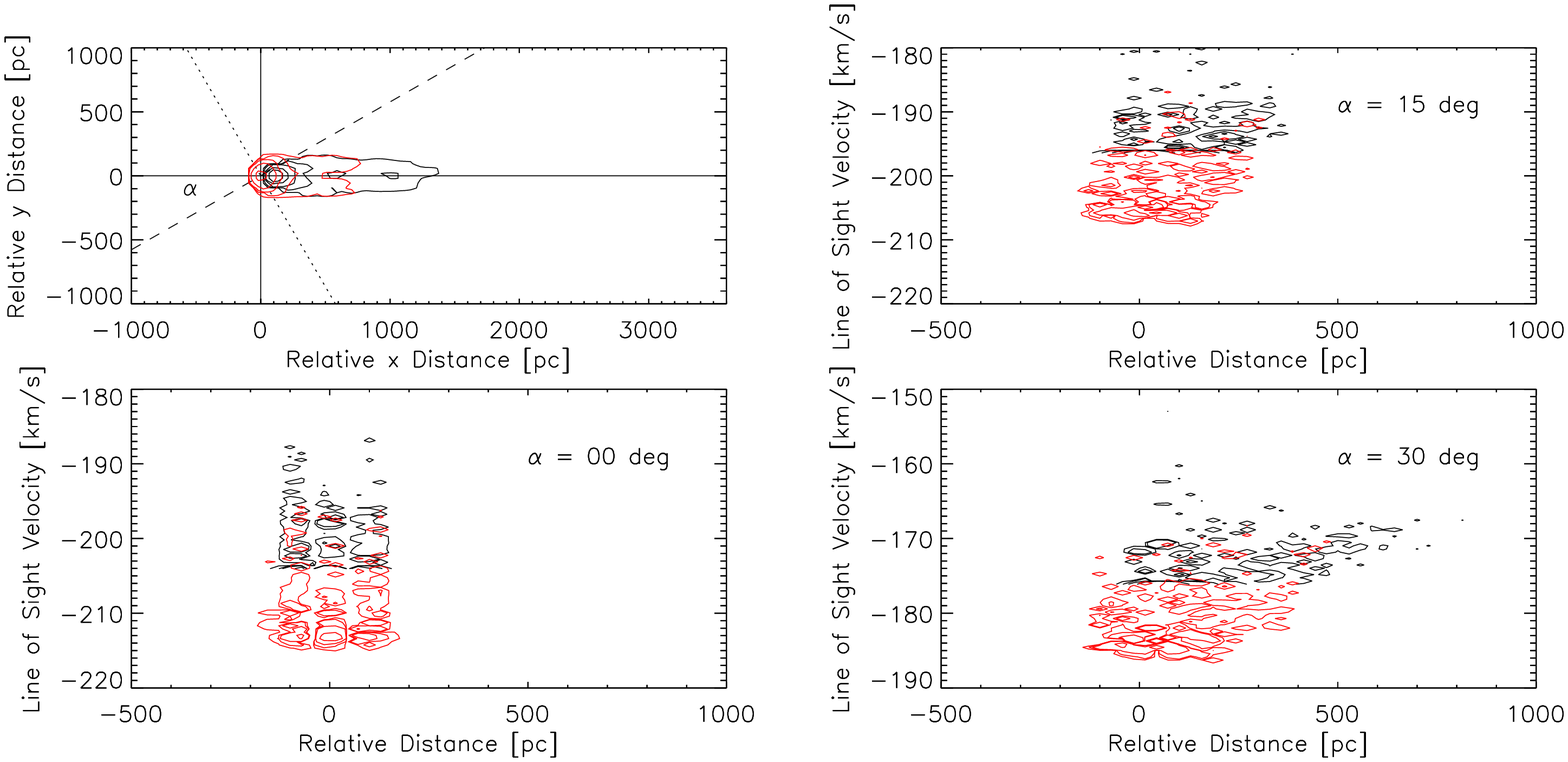}
\caption{As in Fig.~\ref{fig:set2_red} but for {\bf Setup 2-D.}}
\label{fig:set2D_red}
\end{center}
\end{figure*}

\begin{figure*}[htbp]
\begin{center}
	\includegraphics[width=\linewidth]{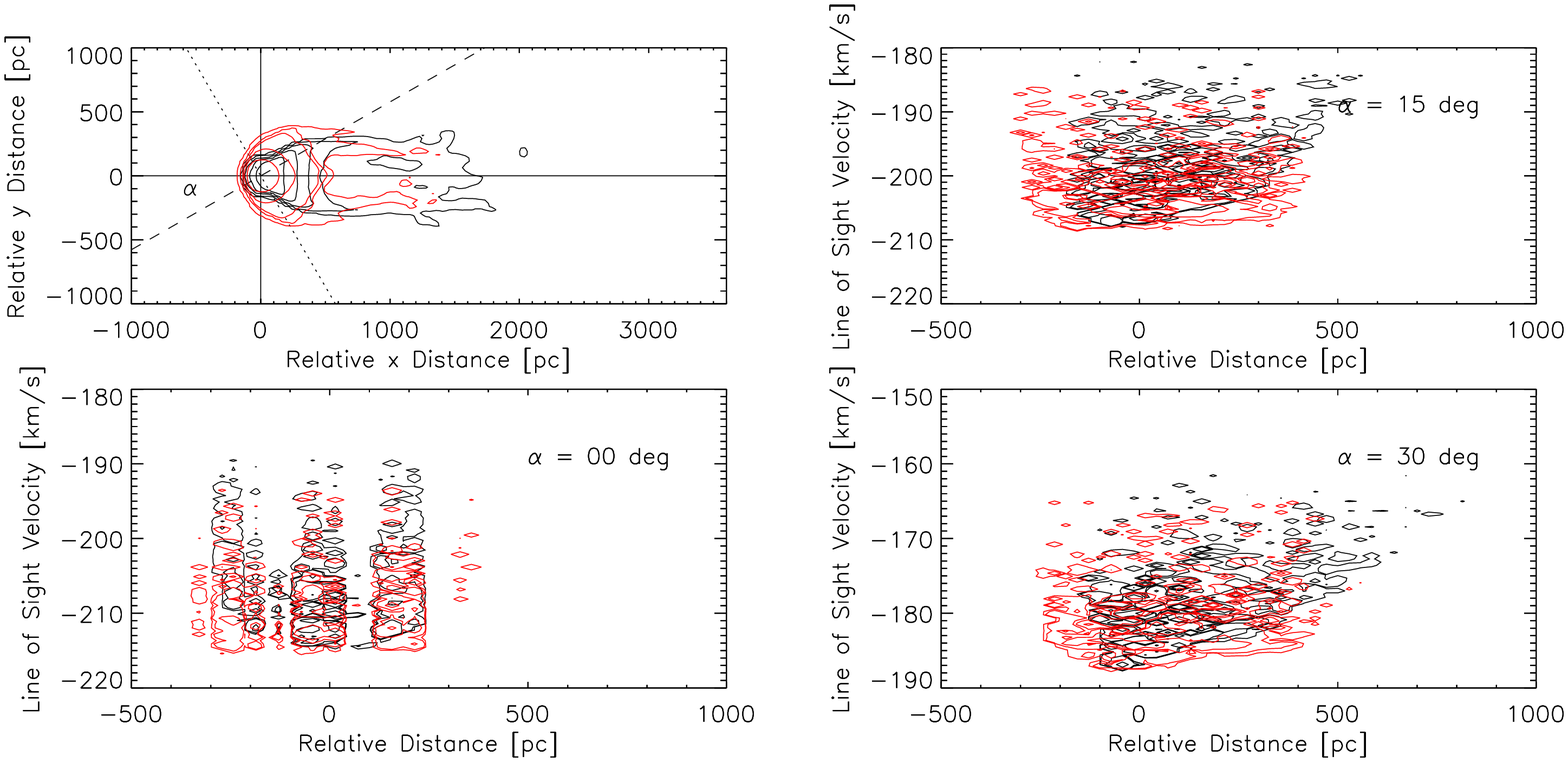}
\caption{As in Fig.~\ref{fig:set2_red} but for {\bf Setup 3.}}
\label{fig:set3_red}
\end{center}
\end{figure*}

\begin{figure*}[htbp]
\begin{center}
	\includegraphics[width=\linewidth]{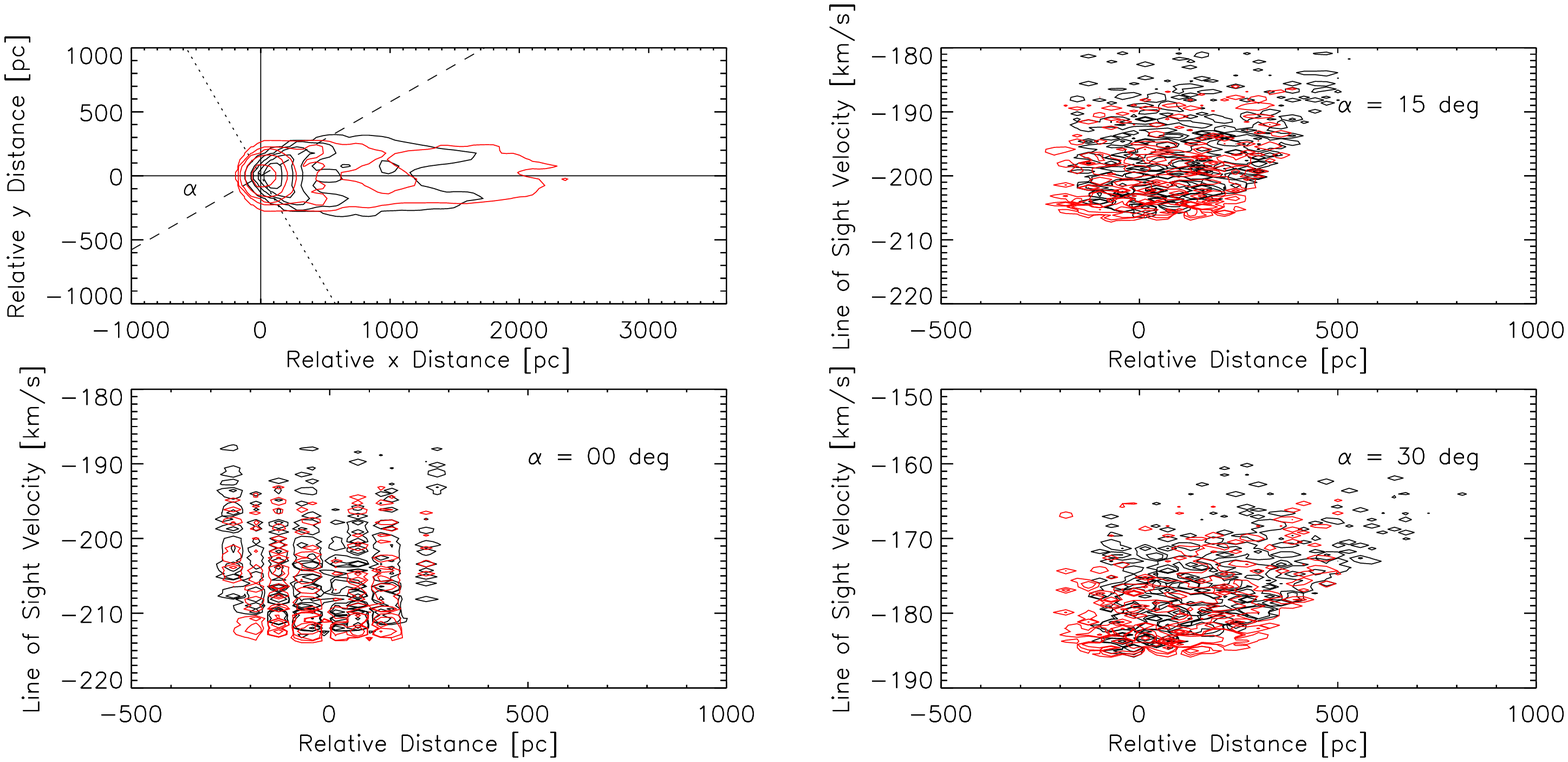}
\caption{As in Fig.~\ref{fig:set2_red} but for {\bf Setup 3-D.}}
\label{fig:set3D_red}
\end{center}
\end{figure*}

\end{appendix}

\end{document}